\documentclass[letterpaper,twocolumn,10pt]{article}

\usepackage[pagebackref=true,breaklinks=true,colorlinks,bookmarks=false]{hyperref}
\usepackage{usenix2019_v3}
\usepackage{tikz}
\usepackage{amsmath,amssymb,amsfonts}
\usepackage{booktabs}
\usepackage{tcolorbox}
\usepackage{subcaption}
\usepackage{color}
\usepackage{multirow}

\newcommand{\model}{\mathcal{M}}

\newcommand{\shadowModel}{\mathcal{M}_{\it shadow}}
\newcommand{\updatedModel}{\mathcal{M'}}

\newcommand{\updatedShadow}{\mathcal{M'}_{\it shadow}}

\newcommand{\update}{\mathcal{D}_{\it update}}
\newcommand{\dataset}{\mathcal{D}}
\newcommand{\probing}{\mathcal{D}_{\it probe}}
\newcommand{\shadowData}{\mathcal{D}_{\it shadow}}
\newcommand{\targetData}{\mathcal{D}_{\it target}}

\newcommand{\shadowAV}{\delta_{\it shadow}}
\newcommand{\attackVector}{\delta}
\newcommand{\attacker}{\mathcal{A}}
\newcommand{\featurevec}{x}
\newcommand{\outputvec}{y}
\newcommand{\labelVec}{\ell}
\newcommand{\distvec}{q}
\newcommand{\UGAN}{CBM-GAN}

\newcommand{\latentVector}{\mu}

\begin{document}
%-------------------------------------------------------------------------------

\date{}

\title{\Large \bf Updates-Leak: Data Set Inference and Reconstruction Attacks in Online Learning}

\author{
{\rm Ahmed Salem}\\
CISPA Helmholtz Center\\
for Information Security
\and
{\rm Apratim Bhattacharya}\\
Max Planck Institute\\
for Informatics
\and
{\rm Michael Backes}\\
CISPA Helmholtz Center\\
for Information Security
\and
{\rm Mario Fritz}\\
CISPA Helmholtz Center\\
for Information Security
\and
{\rm Yang Zhang}\\
CISPA Helmholtz Center\\
for Information Security
} % end author

\maketitle

%-------------------------------------------------------------------------------
\begin{abstract}
%-------------------------------------------------------------------------------
Machine learning (ML) has progressed rapidly during the past decade and the major factor that drives such development is the unprecedented large-scale data. As data generation is a continuous process, this leads to ML model owners updating their models frequently with newly-collected data in an online learning scenario. In consequence, if an ML model is queried with the same set of data samples at two different points in time, it will provide different results.

In this paper, we investigate whether the change in the output of a black-box ML model before and after being updated can leak information of the dataset used to perform the update, namely the updating set. This constitutes a new attack surface against black-box ML models and such information leakage may compromise the intellectual property and data privacy of the ML model owner. We propose four attacks following an encoder-decoder formulation, which allows inferring diverse information of the updating set. Our new attacks are facilitated by state-of-the-art deep learning techniques. In particular, we propose a hybrid generative model ({\UGAN}) that is based on generative adversarial networks (GANs) but includes a reconstructive loss that allows reconstructing accurate samples. Our experiments show that the proposed attacks achieve strong performance.
\end{abstract}

%-------------------------------------------------------------------------------
\section{Introduction}

Machine learning (ML) has progressed rapidly during the past decade.
A key factor that drives the current ML development is the unprecedented large-scale data.
In consequence, collecting high-quality data becomes essential
for building advanced ML models.
Data collection is a continuous process,
which in turn transforms the ML model training into a continuous process as well:
Instead of training an ML model for once and keeping on using it afterwards, 
the model's owner
needs to keep on updating the model with newly-collected data.
As training from scratch is often prohibitive, this is often achieved by \emph{online learning}.
We refer to the dataset used to perform model update as the \emph{updating set}.

In this paper, our main research question is:
\emph{ Can different outputs of an ML model's two versions 
queried with the same set of data samples
leak information of the corresponding updating set?}.
This constitutes a new attack surface against machine learning models.
Information leakage of the updating set 
may compromise the intellectual property and data privacy of the model owner.

We concentrate on the most common ML application -- classification.
More importantly, we target black-box ML models -- 
the most difficult attack setting 
where an adversary does not have access to her target model's parameters
but can only query the model with her data samples and obtain the corresponding prediction results, 
i.e., \emph{posteriors} in the case of classification.
Moreover, we assume the adversary has a local dataset 
from the same distribution as the target model’s training set, 
and the ability to establish the same model as the target model with respect to model architecture.
Finally, we only consider updating sets which contain up to 100 newly collected data samples. 
Note that this is a simplified setting and a step towards real-world setting.

In total, we propose four different attacks in this surface
which can be categorized into two classes,
namely, \emph{single-sample attack class} and \emph{multi-sample attack class}.
The two attacks in the single-sample attack class 
concentrate on a simplified case when the target ML model is updated with one single data sample.
We investigate this case to show 
whether an ML model's two versions' different outputs indeed constitute a valid attack surface.
The two attacks in the multi-sample attack class 
tackle a more general and complex case when the updating set contains multiple data samples.

Among our four attacks,
two (one for each attack class) aim at reconstructing the updating set
which are the first attempts in this direction.
Compared to many previous attacks 
inferring certain properties of a target model's training set~\cite{FJR15,HAP17,GWYGB18},
a dataset reconstruction attack
leads to more severe consequences.

Our experiments show that indeed, the output difference of the same ML model's
two different versions can be exploited to infer information about the updating set.
We detail our contributions as the following.

\smallskip
\noindent\textbf{General Attack Construction.}
Our four attacks follow a general structure, which can be formulated into an encoder-decoder style.
The encoder realized by a multilayer perceptron (MLP) 
takes the difference of the target ML model's outputs, namely \emph{posterior difference}, as its input
while the decoder produces different types of information about the updating set
with respect to different attacks.

To obtain the posterior difference,
we randomly select a fixed set of data samples, namely \emph{probing set},
and probe the target model's two different versions
(the second-version model is obtained by updating the first-version model with an updating set).
Then, we calculate the difference between the two sets of posteriors as the input for our attack's encoder.

\smallskip
\noindent\textbf{Single-Sample Attack Class.}
The single-sample attack class contains two attacks: 
\emph{Single-sample label inference attack} and \emph{single-sample reconstruction attack}.
The first attack predicts the label of the single sample used to update the target model.
We realize the corresponding decoder for the attack by a two-layer MLP.
Our evaluation shows that our attack is able to achieve a strong performance, e.g., 0.96 accuracy 
on the CIFAR-10 dataset~\cite{CIFAR}.

The single-sample reconstruction attack aims at reconstructing the updating sample.
We rely on autoencoder (AE).
In detail, we first train an AE on a different set of data samples.
Then, we transfer the AE's decoder into our attack model as its sample reconstructor.
Experimental results show that 
we can reconstruct the single sample with
a performance gain (with respect to mean squared error) 
of 22\% for the MNIST dataset~\cite{MNIST}, 107.1\% for the CIFAR-10 dataset, 
and 114.7\% for the Insta-NY dataset~\cite{BHPZ17},
over randomly picking a sample affiliated with the same label of the updating sample.

\smallskip
\noindent\textbf{Multi-Sample Attack Class.}
The multi-sample attack class includes
\emph{multi-sample label distribution estimation attack}
and \emph{multi-sample reconstruction attack}.
Multi-sample label distribution estimation attack
estimates the label distribution of the updating set's data samples.
It is a generalization of the label inference attack in the single-sample attack class.
We realize this attack by setting up the attack model's decoder 
as a multilayer perceptron with a fully connected layer and a softmax layer.
Kullback-Leibler divergence (KL-divergence) is adopted as the model's loss function.
Our experiments demonstrate the effectiveness of this attack.
For the CIFAR-10 dataset, when the updating set's cardinality is 100, 
our attack model achieves a 0.00384 KL-divergence
which outperforms random guessing by a factor of 2.5.
Moreover, the accuracy of predicting the most frequent label 
is 0.29 which is almost 3 times higher than random guessing.

Our last attack, namely multi-sample reconstruction attack, 
aims at generating all samples in the updating set.
This is a much more complex attack than the previous ones.
The decoder for this attack is assembled with two components.
The first one learns the data distribution of the updating set samples.
In order to achieve coverage and accuracy of the reconstructed samples, 
we propose a novel hybrid generative model, namely {\UGAN}. 
Different from the standard generative adversarial networks (GANs), 
our Conditional Best of Many GAN ({\UGAN}) introduces a ``Best Match'' loss
which ensures that \textit{each} sample in the updating set is reconstructed \textit{accurately}.
The second component of our decoder relies on machine learning clustering
to group the generated data samples by {\UGAN} into clusters
and take the central sample of each cluster as one final reconstructed sample.
Our evaluation shows that
our approach outperforms all baselines 
when reconstructing the updating set on all MNIST, CIFAR-10, and Insta-NY datasets.
\section{Preliminaries}
\label{section:preli}

In this section, we start by introducing online learning, then present our threat model,
and finally introduce the datasets used in our experiments.

\subsection{Online Learning}

In this paper, we focus on the most common ML task -- classification.
An ML classifier $\model$ is essentially 
a function that maps a data sample $\featurevec \in \mathcal{X}$
to posterior probabilities $\outputvec \in \mathcal{Y}$, i.e.,
$\model: \mathcal{X} \rightarrow \mathcal{Y}$.
Here, $\outputvec \in \mathcal{Y}$ is a vector 
with each entry indicating the probability of $x$ being classified 
to a certain class or affiliated with a certain label.
The sum of all values in $\outputvec$ is 1.
To train an ML model, we need a set of data samples, i.e., training set.
The training process is performed by a certain optimization algorithm, such as ADAM,
following a predefined loss function.

A trained ML model $\model$ can be updated with an updating set denoted by $\update$.
The model update is performed by further training the model 
with the updating set using the same optimization algorithm on the basis of the current model's parameters.
More formally, given an updating set $\update$ 
and a trained ML model $\model$, the updating process $\mathcal{F}_{\it update}$ can be defined as
$\mathcal{F}_{\it update}: \update, \model \rightarrow \updatedModel$
where $\updatedModel$ is the updated version of $\model$.

\subsection{Threat Model}
\label{subsection:threatModel}

For all of our four attacks, we consider an adversary with black-box access to the target model.
This means that the adversary can only query the model with a set of data samples, i.e., her probing set,
and obtain the corresponding posteriors.
This is the most difficult attack setting for the adversary~\cite{SSSS17}.
We also assume that the adversary has a local dataset
which comes from the same distribution as the target model's training set
following previous works~\cite{SSSS17,GWYGB18,SZHBFB19}.
Moreover, we consider the adversary to be able to establish the same ML model
as the target ML model with respect to model architecture.
This can be achieved by
performing model hyperparameter stealing attacks~\cite{WG18,OASF18}.
The adversary needs these two information 
to establish a shadow model 
which mimics the behavior of the target model to derive data for training her attack model 
(see \autoref{section:gap}).
Also, part of the adversary's local dataset will be used as her probing set.
Finally, we assume that the target ML model is updated only with new data, 
i.e., the updating set and the training set are disjoint.

We later show in~\autoref{section:further_analysis} that the two assumptions,
i.e., the adversary's
knowledge of the target model's architecture and her possession of a dataset 
from the same distribution as the target model's training set,
can be further relaxed.

\begin{figure*}[!t]
\centering
\includegraphics[width=1.8\columnwidth]{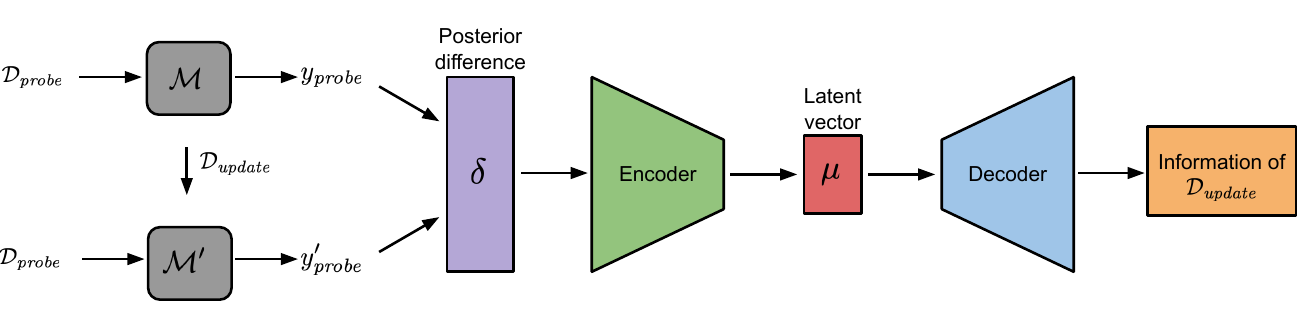}
\caption{A schematic view of the general attack pipeline.}
\label{fig:attack_overview}
\end{figure*}

\subsection{Datasets Description}

For our experimental evaluation, we use three datasets: MNIST, CIFAR-10, and Insta-NY. 
Both MNIST and CIFAR-10 are benchmark datasets for various ML security and privacy tasks. 
MNIST is a 10-class image dataset, it consists of 70,000 28$\times$28 grey-scale images. 
Each image contains in its center a handwritten digit. 
Images in MNIST are equally distributed over 10 classes.
CIFAR-10 contains 60,000 32$\times$32 color images. 
Similar to MNIST, CIFAR-10 is also a 10-class balanced dataset.
Insta-NY~\cite{BHPZ17} contains a sample of Instagram users' location check-in data in New York.
Each check-in represents a user visiting a certain location at a certain time.
Each location is affiliated with a category. In total, there are eight different categories.
Our ML task for Insta-NY is to predict each location's category.
We use the number of check-ins happened at each location 
in each hour on a weekly base as the location's feature vector.
We further filter out locations with less than 50 check-ins,
in total, we have 19,215 locations for the dataset.
In~\autoref{section:further_analysis}, 
we further use Insta-LA~\cite{BHPZ17}
which contains the check-in data from Los Angeles 
for our threat model relaxation experiments.
\section{General Attack Pipeline}
\label{section:gap}

Our general attack pipeline contains three phases.
In the first phase, the adversary generates her attack input, i.e., posterior difference.
In the second phase, our encoder transforms the posterior difference into a latent vector.
In the last phase, the decoder decodes the latent vector to produce different information of the updating set 
with respect to different attacks.
\autoref{fig:attack_overview} provides a schematic view of our attack pipeline.

In this section, we provide a general introduction for each phase of our attack pipeline.
In the end, we present our strategy of deriving data to train our attack models.

\subsection{Attack Input}
\label{subsection:attackInput}

Recall that we aim at investigating
the information leaked from posterior difference
of a model's two versions when queried with the same set of data samples.
To create this posterior difference,
the adversary first needs to pick a set of data samples as her probing set, denoted by $\probing$.
In this work, the adversary picks a random sample of data samples (from her local dataset) to form $\probing$.
Choosing or crafting~\cite{OASF18} a specific set of data samples as the probing set 
may further improve attack efficiency,
we leave this as a future work.
Next, the adversary queries the target ML model $\model$ with all samples in $\probing$ 
and concatenates the received outputs to form a vector $\outputvec_{\it probe}$.
Then, she probes the updated model $\updatedModel$ with samples in $\probing$
and creates a vector $\outputvec'_{\it probe}$ accordingly.
In the end, she sets the posterior difference, denoted by $\attackVector$, to the difference of both outputs:
\[
\attackVector = \outputvec_{\it probe} - \outputvec'_{\it probe}
\]
Note that the dimension of $\attackVector$ is the product of
$\probing$'s cardinality and the number of classes of the target dataset.
For this paper, both CIFAR-10 and MNIST are 10-class datasets, while Insta-NY is an 8-class dataset.
As our probing set always contains 100 data samples, 
this indicates the dimension of $\attackVector$ is 1,000 for CIFAR-10 and MNIST, and 800 for Insta-NY.

\subsection{Encoder Design}
\label{subsection:encStruc}

All our attacks share the same encoder structure,
we model it with a multilayer perceptron.
The number of layers inside the encoder depends on the dimension of $\attackVector$: 
Longer $\attackVector$ requires more layers in the encoder. 
As our $\attackVector$ is a 1,000-dimension vector for the MNIST and CIFAR-10 datasets, 
and 800-dimension vector for the Insta-NY dataset,
we use two fully connected layers in the encoder.
The first layer transforms $\attackVector$ to a 128-dimension vector
and the second layer further reduces the dimension to 64.
The concrete architecture of our encoder is presented in~\autoref{section:encArc}.

\subsection{Decoder Structure}

Our four attacks aim at inferring different information of $\update$,
ranging from sample labels to the updating set itself.
Thus, we construct different decoders for different attacks with different techniques. 
The details of these decoders will be presented in the following sections.

\subsection{Shadow Model}
\label{subsection:shadowModel}

Our encoder and decoder need to be trained jointly in a supervised manner.
This indicates that we need ground truth data for model training.
Due to our minimal assumptions, the adversary cannot get the ground truth from the target model.
To solve this problem, we rely on shadow models following previous works~\cite{SSSS17,GWYGB18,SZHBFB19}.
A shadow model is designed to mimic the target model.
By controlling the training process of the shadow model, 
the adversary can derive the ground truth data needed to train her attack models.

As presented in \autoref{section:preli}, 
our adversary knows (1) the architecture of the target model 
and (2) a dataset coming from the same distribution as the target dataset. 
To build a shadow model $\shadowModel$, 
the adversary first establishes an ML model with the same structure as the target model.
Then, she gets a shadow dataset $\shadowData$ from her local dataset (the rest is used as $\probing$) 
and splits it into two parts: 
Shadow training set $\shadowData^{\it train}$ and shadow updating set $\shadowData^{\it update}$.
$\shadowData^{\it train}$ is used to train the shadow model
while $\shadowData^{\it update}$ is further split to $m$ datasets: $\shadowData^{\it update^1} \cdots \shadowData^{\it update^m}$.
The number of samples in each of the $m$ datasets depends on the attack.
For instance, our single-sample class attacks require each dataset containing a single sample.
The adversary then generates $m$ shadow updated models $\updatedShadow^1  \cdots \updatedShadow^m$ 
by updating the shadow model $\shadowModel$ with $m$ shadow updating sets in parallel.

The adversary, in the end, probes the shadow and updated shadow models with her probing set $\dataset_{\it probe}$,
and calculates the shadow posterior difference $\shadowAV^1 \cdots \shadowAV^m$.
Together with the corresponding shadow updating set's ground truth information (depending on the attack),
the training data for her attack model is derived.

More generally, the training set for each of our attack models contains $m$ samples 
corresponding to $\shadowData^{\it update^1} \cdots \shadowData^{\it update^m}$.
In all our experiments, we set $m$ to 10,000.
In addition, we create 1,000 updated models for the target model, 
this means the testing set for each attack model contains 1,000 samples,
corresponding to $\targetData^{\it update^1} \cdots \targetData^{\it update^{1,000}}$.
\section{Single-sample Attacks}
\label{section:ssa}

In this section, we concentrate on the case when an ML model is updated with a single sample.
This is a simplified attack scenario and we aim to examine the possibility of using posterior difference 
to infer information about the updating set.
We start by introducing the single-sample label inference attack,
then, present the single-sample reconstruction attack.

\subsection{Single-sample Label Inference Attack}
\label{subsection:LPAttack}

\noindent\textbf{Attack Definition.}
Our single-sample label inference attack takes the posterior difference 
as the input and outputs the label of the single updating sample.
More formally, given a posterior difference $\attackVector$, 
our single-sample label inference attack is defined as follows:
\[
\attacker_{\it LI}: \attackVector \mapsto \labelVec
\]
where $\labelVec$ is a vector 
with each entry representing
the probability of the updating sample affiliated with a certain label.

\smallskip
\noindent\textbf{Methodology.}
To recap, the general construction of the attack model consists of an MLP-based encoder 
which takes the posterior difference as its input and outputs a latent vector $\latentVector$.
For this attack, 
the adversary constructs her decoder also with an MLP
which is assembled with a fully connected layer and a softmax layer 
to transform the latent vector to the corresponding updating sample's label.
The concrete architecture of our $\attacker_{\it LI}$'s decoder is presented in~\autoref{section:SLIArc}.

To obtain the data for training $\attacker_{\it LI}$,
the adversary generates ground truth data by creating a shadow model 
as introduced in \autoref{section:gap} while setting the shadow updating set's cardinality to 1.
Then, the adversary trains her attack model $\attacker_{\it LI}$ with a cross-entropy loss. Our loss function is,
\begin{align*}
    \mathcal{L}_{\it CE} = \sum\limits_{i}  \labelVec_{i} \log(\hat{\labelVec}_{i})
\end{align*}
where $\labelVec_{i}$ is the true probability of label $i$ 
and $\hat{\labelVec}_{i}$ is our predicted probability of label $i$.
The optimization is performed by the ADAM optimizer.

To perform the label inference attack, 
the adversary constructs the posterior difference as introduced in \autoref{section:gap}, 
then feeds it to the attack model $\attacker_{\it LI}$ to obtain the label.

\smallskip
\noindent\textbf{Experimental Setup.}
We evaluate the performance of our single-sample label inference attack 
using the MNIST, CIFAR-10, and Insta-NY datasets.
First, we split each dataset into three disjoint datasets: 
The target dataset $\targetData$, the shadow dataset $\shadowData$, and the probing dataset $\probing$.
As mentioned before, $\probing$ contains 100 data samples.
We then split $\shadowData$ to $\shadowData^{\it train}$ and $\shadowData^{\it update}$ 
to train the shadow model as well as updating it (see~\autoref{section:gap}).
The same process is applied to train and update the target model with $\targetData$.
As mentioned in~\autoref{section:gap}, 
we build 10,000 and 1,000 updated models for shadow and target models, respectively.
This means the training and testing sets for our attack model contain 10,000 and 1,000 samples, respectively.

We use convolutional neural network (CNN) to build shadow and target models
for both CIFAR-10 and MNIST datasets, 
and a multilayer perceptron (MLP) for the Insta-NY dataset.
The CIFAR-10 model consists of two convolutional layers, 
one max pooling layer, three fully connected layers, and a softmax layer.
The MNIST model consists of two convolutional layers, two fully connected layers, and a softmax layer.
Finally, the Insta-NY model consists of three fully connected layers and a softmax layer.
The concrete architectures of the models are presented in~\autoref{section:targetArc}.

All shadow and target models' training sets 
contain 10,000 images for CIFAR-10 and MNIST, and 5,000 samples for Insta-NY.
We train the CIFAR-10, MNIST and Insta-NY models for 50, 25, and 25 epochs, respectively, 
with a batch size of 64.
To create an updated ML model, we perform a single-epoch training.
Finally, we adopt accuracy to measure the performance of the attack.
All of our experiments are implemented using Pytorch~\cite{PyTorch}.
For reproducibility purposes, our code will be made available. 

\begin{figure}[!t]
\centering
\includegraphics[width=0.8\columnwidth]{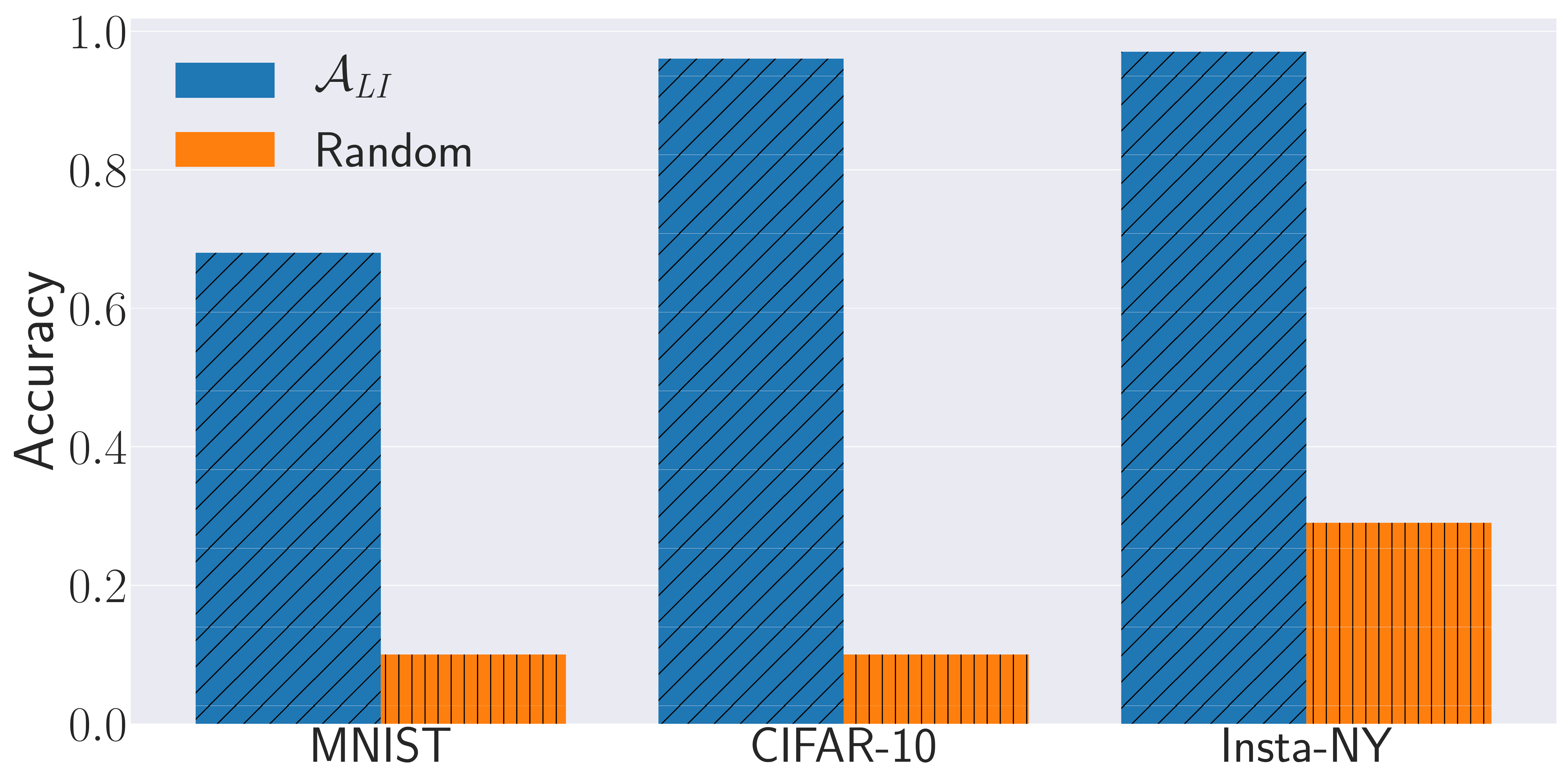}
\caption{
[Higher is better]
Performance of the single-sample label inference attack ($\attacker_{\it LI}$) 
on MNIST, CIFAR-10, and Insta-NY datasets together with the baseline model.
Accuracy is adopted as the evaluation metric.
}
\label{fig:liattack}
\end{figure}

\smallskip
\noindent\textbf{Results.}
\autoref{fig:liattack} depicts the experimental results.
As we can see, $\attacker_{\it LI}$ achieves a strong performance 
with an accuracy of 0.97 on the Insta-NY dataset, 0.96 on the CIFAR-10 dataset,
and 0.68 on the MNIST dataset.
Moreover, our attack significantly outperforms the baseline model, namely Random, 
which simply guesses a label over all possible labels.
As both CIFAR-10 and MNIST contain 10 balanced classes, the baseline model's result is approximately 10\%.
For the Insta-NY dataset, since it is not balanced, 
we randomly sample a label for each sample to calculate the baseline which results in approximately 29\% accuracy.
Our evaluation shows that 
the different outputs of an ML model's two versions indeed leak information of the corresponding updating set.

\subsection{Single-sample Reconstruction Attack}
\label{subsection:SSRA}

\begin{figure}[!t]
\centering
\includegraphics[width=1\columnwidth]{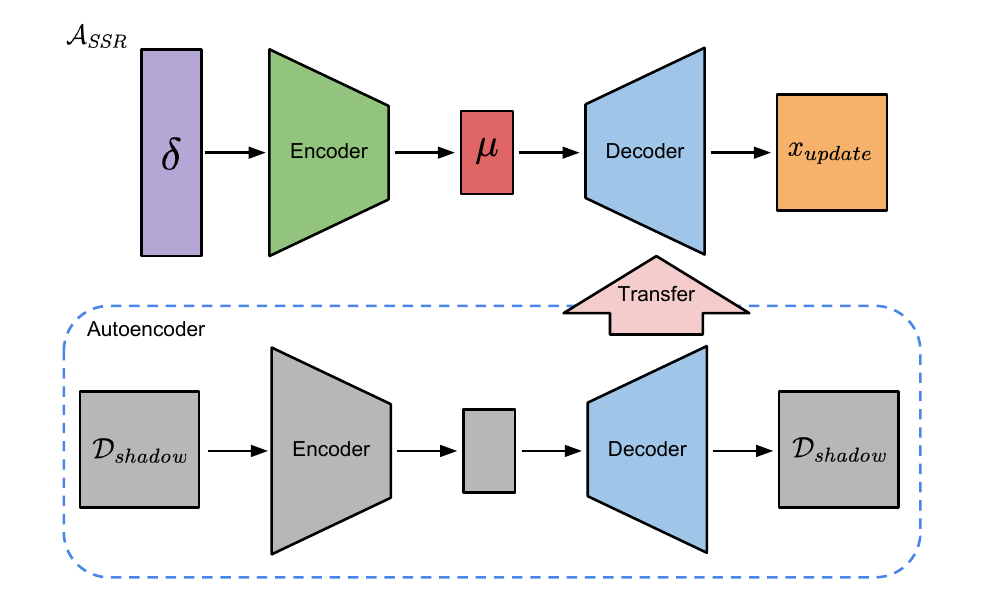}
\caption{Methodology of the single-sample reconstruction attack ($\attacker_{\it SSR}$).}
\label{fig:ssr_overview}
\end{figure}

\begin{figure*}[!t]
\centering
\begin{subfigure}{0.65\columnwidth}
   \includegraphics[width=\columnwidth]{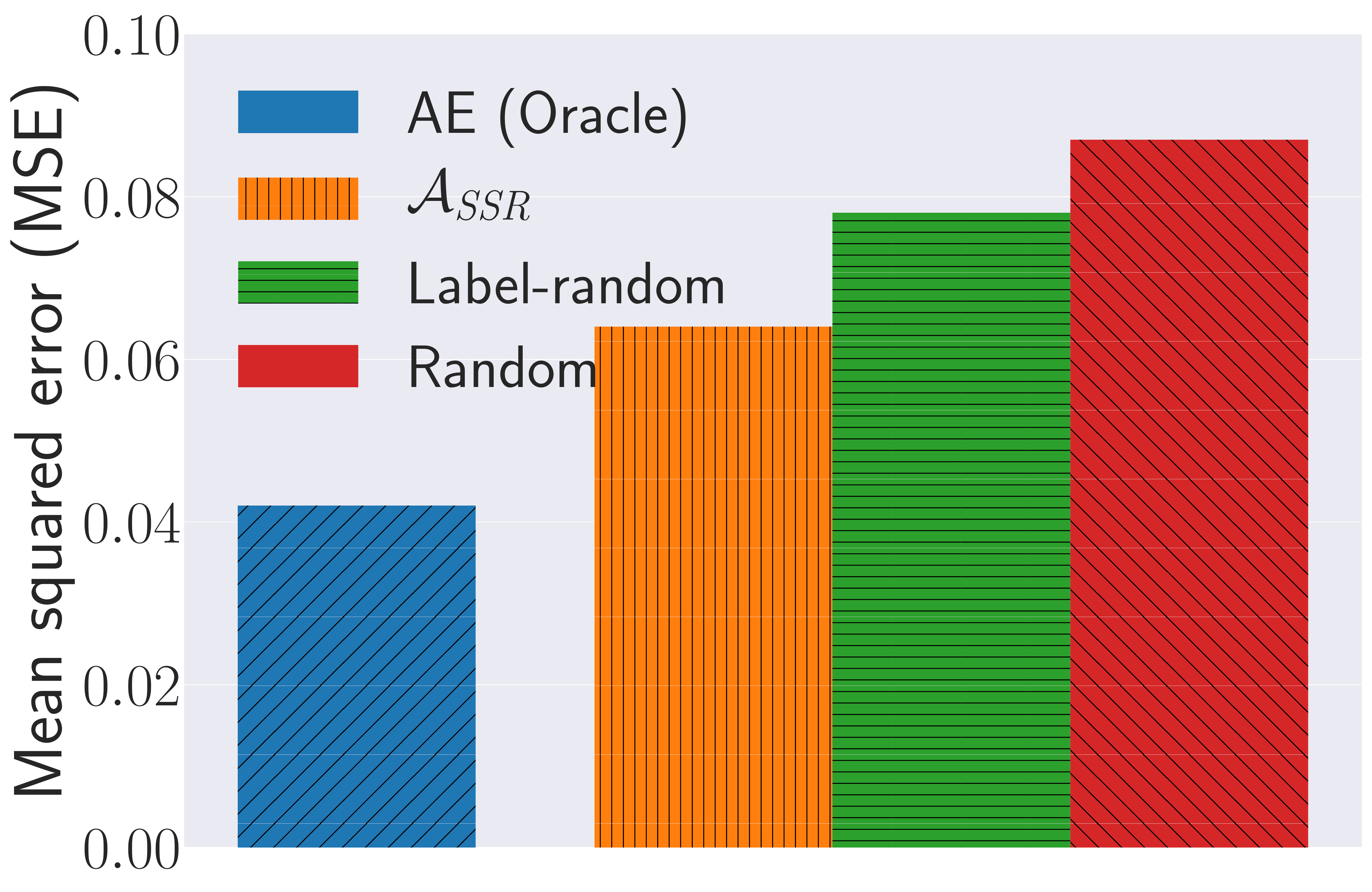}
   \caption{MNIST}
   \label{fig:spr_mse_mnist} 
\end{subfigure}
\begin{subfigure}{0.65\columnwidth}
   \includegraphics[width=\columnwidth]{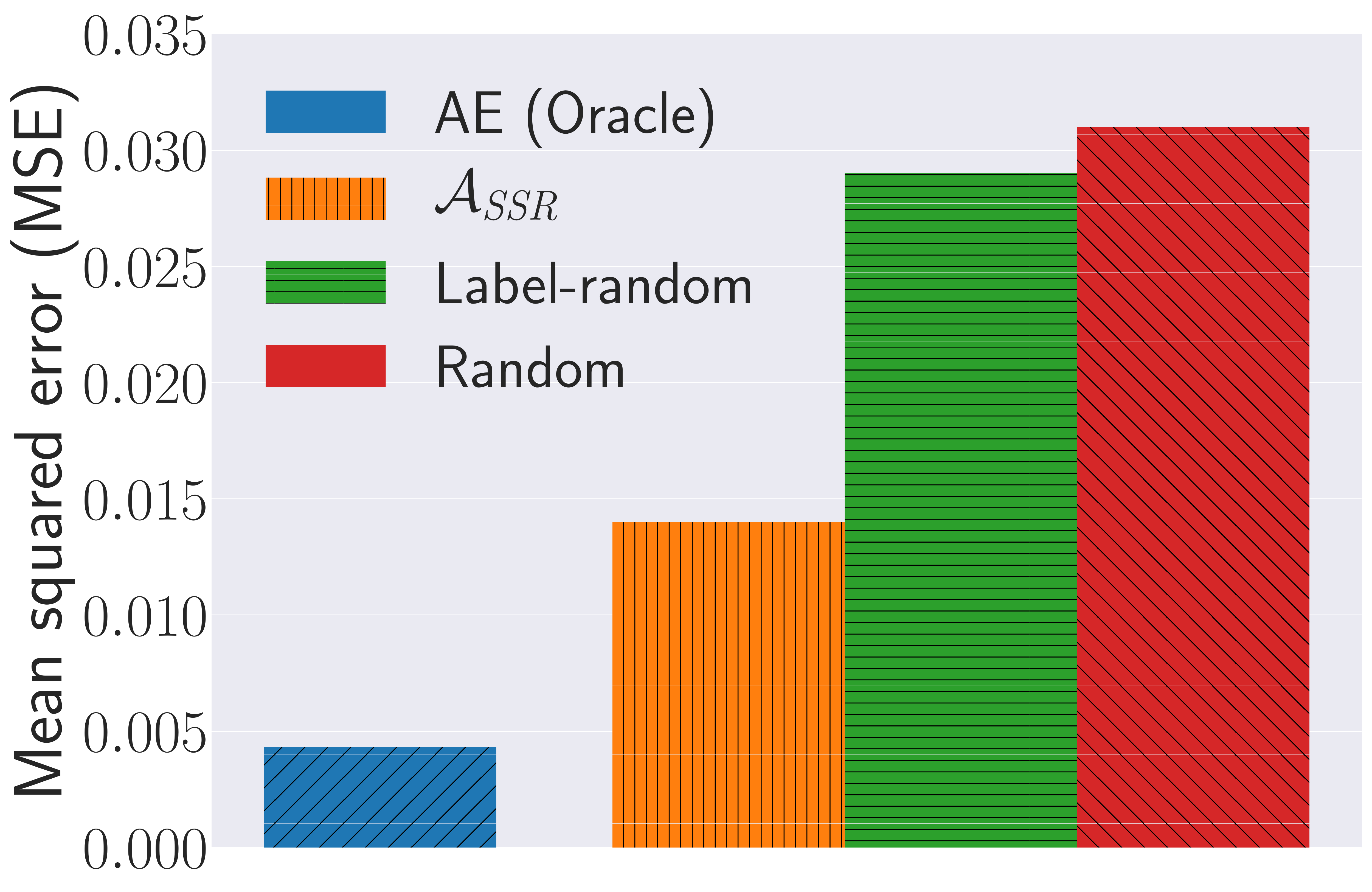}
   \caption{CIFAR-10}
   \label{fig:spr_mse_cifar10} 
\end{subfigure}
\begin{subfigure}{0.65\columnwidth}
   \includegraphics[width=\columnwidth]{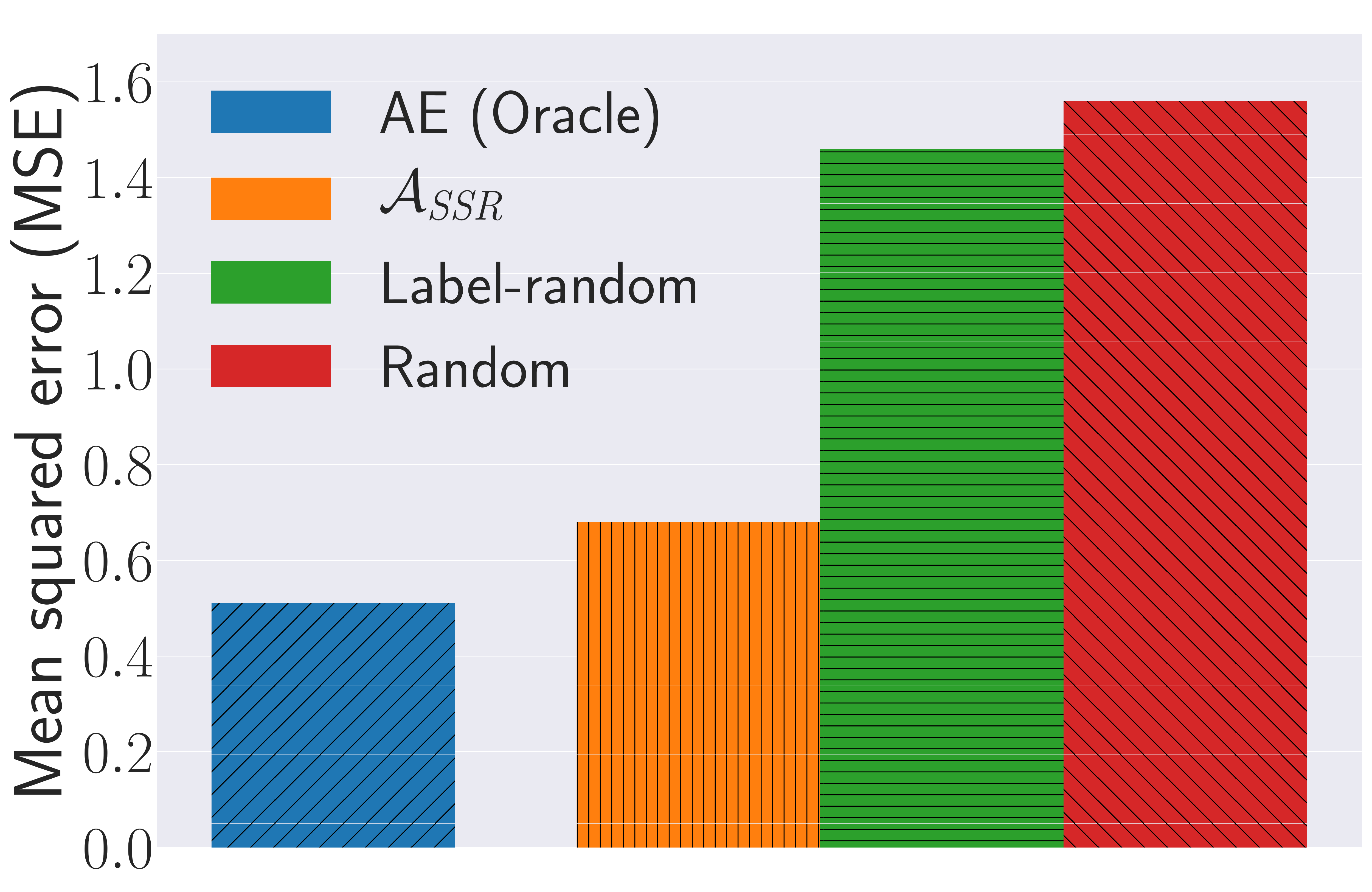}
   \caption{Insta-NY}
   \label{fig:spr_mse_location} 
\end{subfigure}
\caption{
[Lower is better]
Performance of the single-sample reconstruction attack ($\attacker_{\it SSR}$) 
together with autoencoder and two baseline models.
Mean squared error is adopted as the evaluation metric.
Autoencoder (AE) serves as an oracle as the adversary cannot use it for her attack.} 
\label{fig:spr_mse}
\end{figure*}
\begin{figure}[!t]
\centering
\includegraphics[width=0.8\columnwidth]{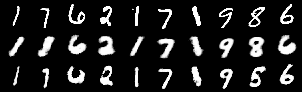}
\caption{Visualization of some generated samples from the single-sample reconstruction attack ($\attacker_{\it SSR}$) 
on the MNIST dataset.
Samples are fair random draws, not cherry-picked.
The first row shows the original samples.
The second row shows the reconstructed samples by $\attacker_{\it SSR}$.
The third shows row the reconstructed samples by autoencoder, i.e., the upper bound of our reconstruction attack.} 
\label{fig:singlePointMNIST}
\end{figure}

\noindent\textbf{Attack Definition.}
Our single-sample reconstruction attack
takes one step further to construct the data sample used to update the model.
Formally, given a posterior difference $\attackVector$, 
our single-sample reconstruction attack, denoted by $\attacker_{\it SSR}$, is defined as follows:
\[
\attacker_{\it SSR}: \attackVector \mapsto  \featurevec_{\it update}
\]
where $\featurevec_{\it update}$ denotes the sample used to update the model ($\update = \{\featurevec_{\it update}\}$).

\smallskip
\noindent\textbf{Methodology.}
Reconstructing a data sample is a much more complex task than predicting the sample's label.
To tackle this problem, we need an ML model which is able to generate a data sample in the complex space.
To this end, we rely on autoencoder (AE).

Autoencoder is assembled with an encoder and a decoder.
Different from our attacks, 
AE's goal is to learn an efficient encoding for a data sample:
Its encoder encodes a sample into a latent vector
and its decoder tries to decode the latent vector to reconstruct the same sample.
This indicates AE's decoder itself is a data sample reconstructor.
For our attack, we first train an AE, 
then transfer the AE's decoder to our attack model as the initialization of the attack's decoder.
\autoref{fig:ssr_overview} provides an overview of the attack methodology.
The concrete architectures of our AEs' encoders and decoders are presented in~\autoref{section:SSRArc}.

After the autoencoder is trained, 
the adversary takes its decoder and appends it to her attack model's encoder.
To establish the link,
the adversary adds an additional fully connected layer to its encoder
which transforms the dimensions of the latent vector $\latentVector$ 
to the same dimension as $\latentVector_{\it AE}$.

We divide the attack model training process into two phases.
In the first phase, the adversary uses her shadow dataset to train an AE 
with the previously mentioned model architecture.
In the second phase, she follows the same procedure for single-sample label inference attack
to train her attack model.
Note that the decoder from AE here serves as 
the initialization of the decoder,
this means it will be further trained together with the attack model's encoder.
To train both autoencoder and our attack model,
we use mean squared error (MSE) as the loss function. Our objective is,
\begin{align*}
    \mathcal{L}_{\it MSE} = \lVert \hat{\featurevec}_{\it update} - \featurevec_{\it update} \lVert^{2}_{2}
\end{align*}
where $\hat{\featurevec}_{\it update}$ is our predicted data sample. 
We again adopt ADAM as the optimizer.

\smallskip
\noindent\textbf{Experimental Setup.}
We use the same experimental setup 
as the previous attack (see~\autoref{subsection:LPAttack}) except for the evaluation metric.
In detail, we adopt MSE to measure our attack's performance instead of accuracy.

We construct two baseline models, namely Label-random and Random.
Both of these baseline models take a random data sample from
the adversary's shadow dataset.
The difference is that the Label-random baseline picks a sample
within the same class as the target updating sample,
while the Random baseline 
takes a random data sample from the whole shadow dataset of the adversary.
The Label-random baseline can be implemented 
by first performing our single-sample label inference attack
to learn the label of the data sample
and then picking a random sample affiliated with the same label.

\smallskip
\noindent\textbf{Results.}
First, our single-sample reconstruction attack achieves a promising performance.
As shown in \autoref{fig:spr_mse},
our attack on the MNIST dataset
outperforms the Random baseline by 36\%
and more importantly, outperforms the Label-random baseline by 22\%.
Similarly, for the CIFAR-10 and Insta-NY datasets, our attack achieves an MSE of 0.014 and 0.68
which is significantly better than the two baseline models, 
i.e., it outperforms the Label-random (Random) baselines by a factor of 2.1 (2.2) and 2.1 (2.3), respectively.
The difference between our attack's performance gain over the baseline models 
on the MNIST and on the other datasets is expected 
as the MNIST dataset is more homogeneous compared to the other two.
In other words, the chance of picking a random data sample similar to the updating sample
is much higher in the MNIST dataset than in the other datasets.

Secondly, we compare our attack's performance against the results of the autoencoder 
for sample reconstruction.
Note that AE takes the original data sample as input 
and outputs the reconstructed one, thus it
is considered as an oracle, since the adversary does not have access to the original updating sample.
Here, we just use AE's result to show the best possible result for our attack.
From \autoref{fig:spr_mse}, we observe that AE achieves 0.042, 0.0043, 
and 0.51 MSE for the MNIST, CIFAR-10, and Insta-NY datasets, respectively,
which indeed outperforms our attack.
However, our attack still has a comparable performance.

Finally, \autoref{fig:singlePointMNIST} visualizes 
some randomly sampled reconstructed images by our attack on MNIST.
The first row depicts the original images used to update the models 
and the second row shows the result of our attack.
As we can see, 
our attack is able to reconstruct images that are visually similar to the original sample
with respect to rotation and shape.
We also show the result of AE in the third row in \autoref{fig:singlePointMNIST} 
which as mentioned before, is the upper bound for our attack.
The results from~\autoref{fig:spr_mse} and~\autoref{fig:singlePointMNIST} 
demonstrate
the strong performance of our attack.
\section{Multi-sample Attacks}
\label{section:msa}

\begin{figure*}[!t]
\centering
\begin{subfigure}{0.52\columnwidth}
   \includegraphics[width=\linewidth]{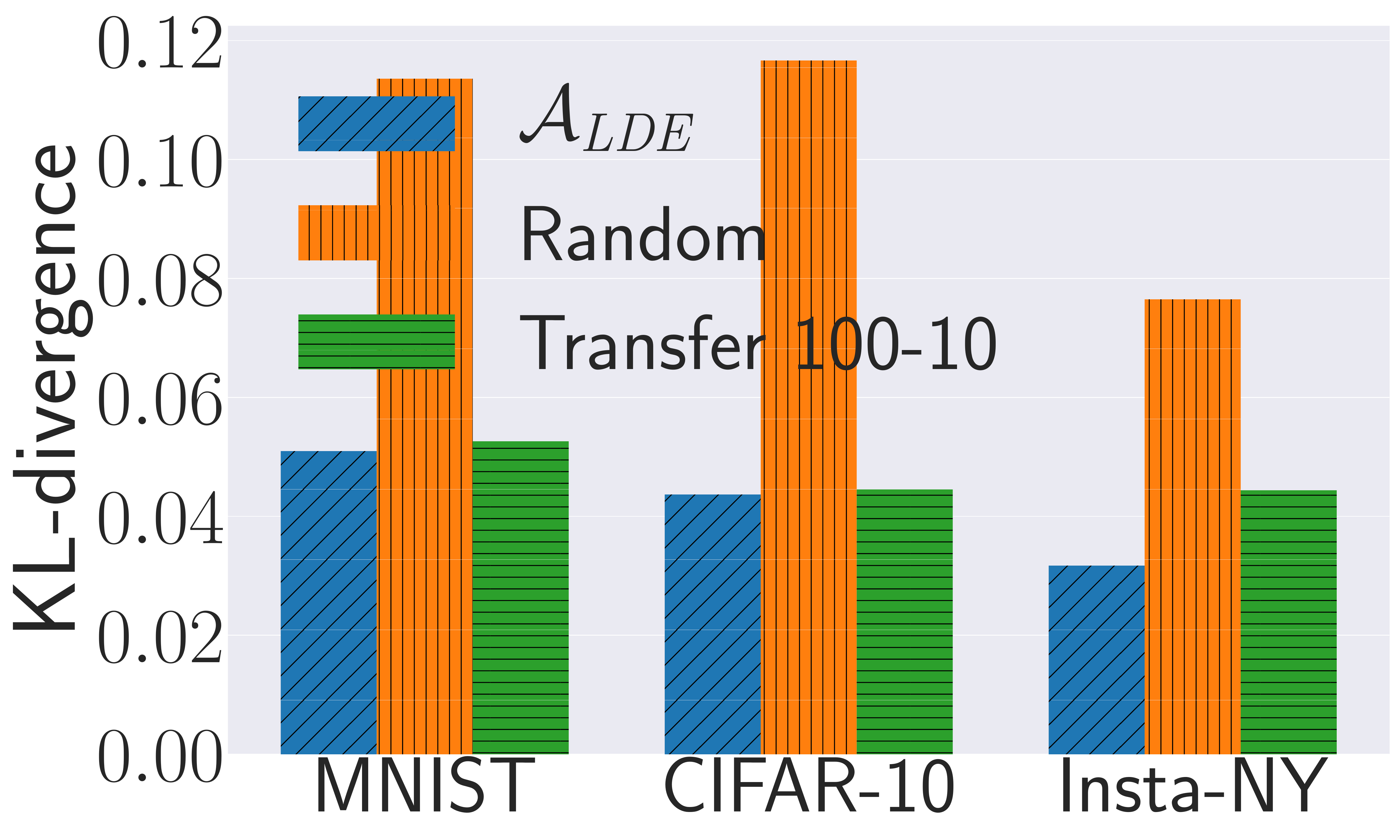}
   \caption{KL-divergence (10 samples)}
   \label{fig:lde_kl10} 
\end{subfigure}
\begin{subfigure}{0.51\columnwidth}
   \includegraphics[width=\linewidth]{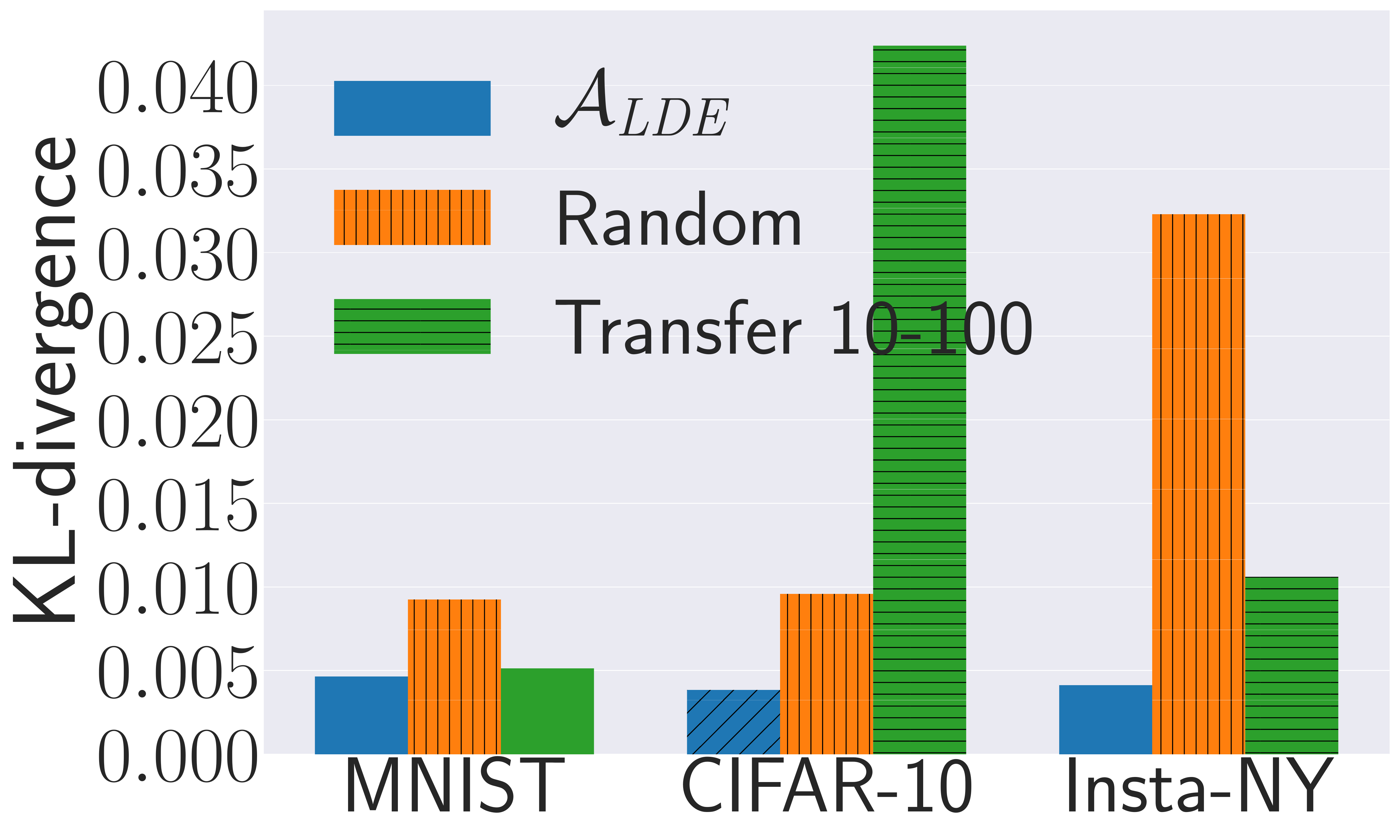}
   \caption{KL-divergence (100 samples)}
   \label{fig:lde_kl100} 
\end{subfigure}
\begin{subfigure}{0.51\columnwidth}
   \includegraphics[width=\linewidth]{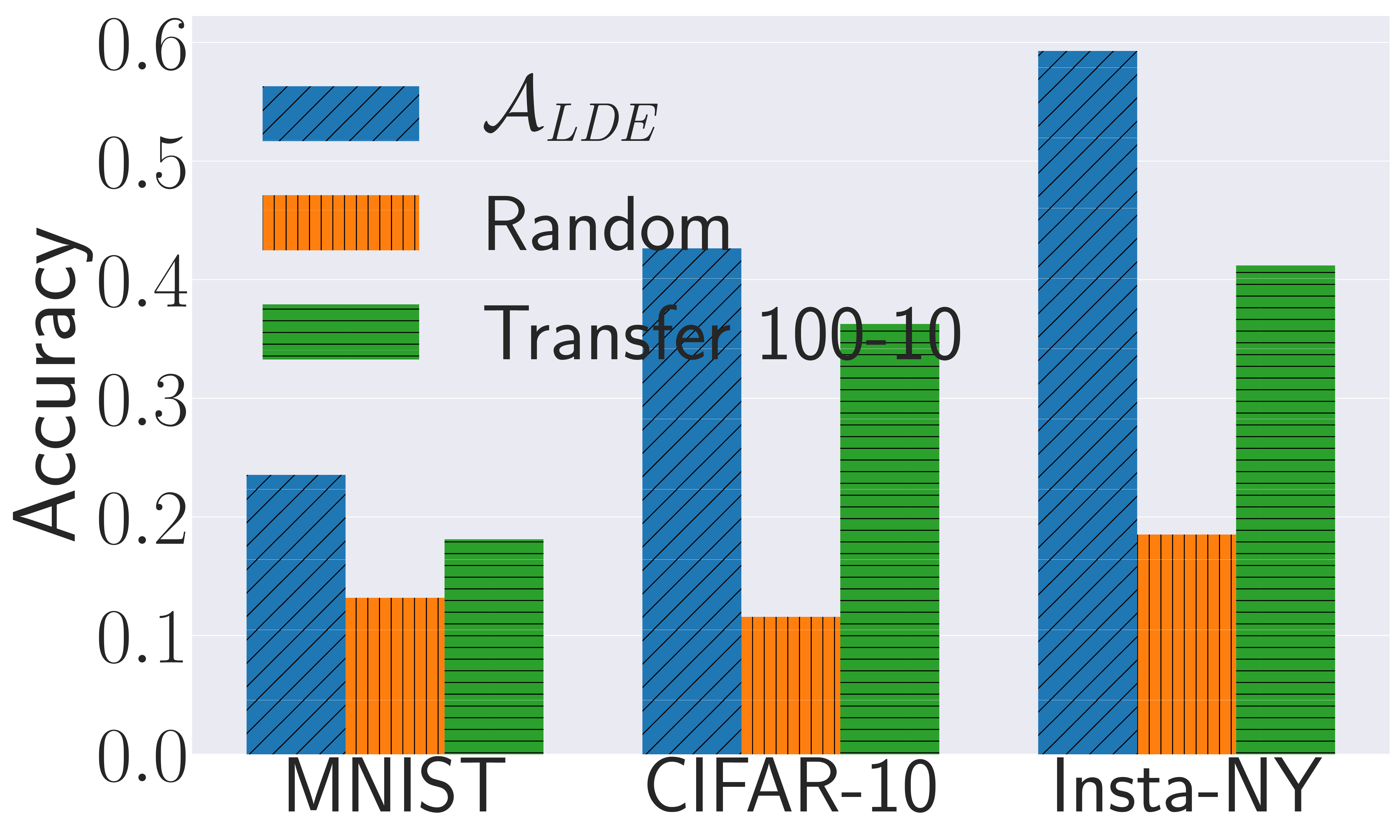}
   \caption{Accuracy (10 samples)}
   \label{fig:lde_accuracy10} 
\end{subfigure}
\begin{subfigure}{0.51\columnwidth}
   \includegraphics[width=\linewidth]{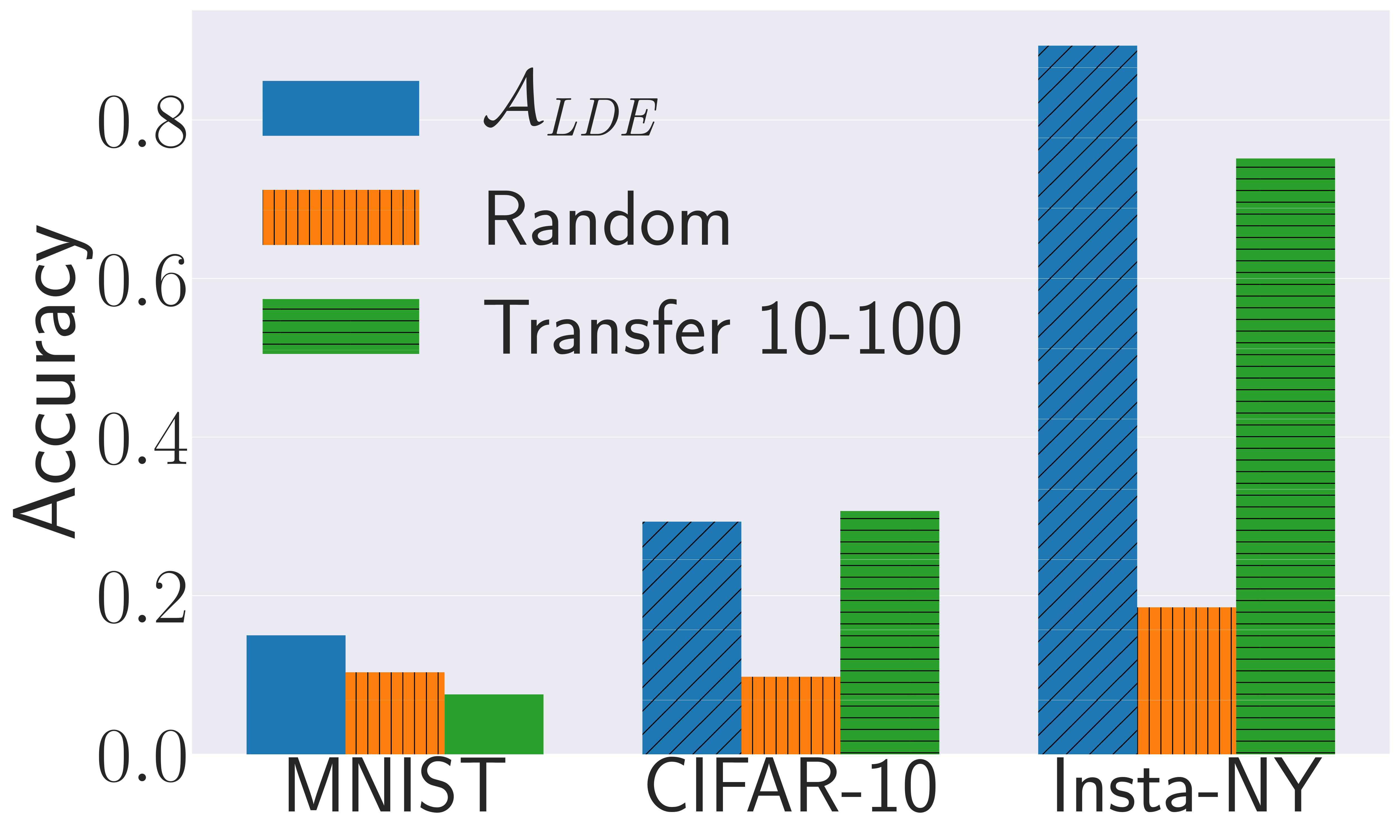}
   \caption{Accuracy (100 samples)}
   \label{fig:lde_accuracy100} 
\end{subfigure}
\caption{
[Lower is better for (a) and (b), higher is better for (c) and (d)]
Performance of the multi-sample label distribution estimation attack ($\attacker_{\it LDE}$) 
together with the baseline model and transfer attack.
KL-divergence and accuracy are adopted as the evaluation metric.
Accuracy here is used to measure the prediction of the most frequent label over samples in the updating set.
Transfer 10-100 means each of the training sample for the attack model 
corresponds to an updating set containing 10 data samples
and each of the testing sample for the attack model 
corresponds to an updating set containing 100 data samples.
} \label{fig:lde}
\end{figure*}

After demonstrating the effectiveness of our attacks against the updating set with a single sample, 
we now focus on a more general attack scenario 
where the updating set contains multiple data samples that are never seen during the training.
We introduce two attacks in the multi-sample attack class:
Multi-sample label distribution estimation attack and multi-sample reconstruction attack.

\subsection{Multi-sample Label Distribution Estimation Attack}
\label{section:LDEAttack}

\noindent\textbf{Attack Definition.}
Our first attack in the multi-label attack class
aims at estimating the label distribution of the updating set's samples.
It can be considered as a generalization 
of the label inference attack in the single-sample attack class.
Formally, the attack is defined as:
\[
\attacker_{\it LDE}: \attackVector \mapsto  \distvec
\]
where $\distvec$ as a vector denotes the distribution of labels 
over all classes for samples in the updating set.

\smallskip
\noindent\textbf{Methodology.}
The adversary uses the same encoder structure as presented in \autoref{section:gap}
and the same decoder structure of the label inference attack (\autoref{subsection:LPAttack}).
Since the label distribution estimation attack 
estimates a probability vector $\distvec$ instead of performing classification,
we use Kullback-–Leibler divergence (KL-divergence) as our objective function:
\begin{align*}
    \mathcal{L}_{\it KL} = \sum\limits_{i} (\hat{\distvec}_{\labelVec})_{i} \log \frac{(\hat{\distvec}_{\labelVec})_{i}}{(\distvec_{\labelVec})_{i}}
\end{align*}
where $\hat{\distvec}_{\labelVec}$ and $\distvec_{\labelVec}$ represent 
our attack's estimated label distribution and the target label distribution, respectively, 
and $(\distvec_{\labelVec})_i$ corresponds to the $i$th label.

To train the attack model $\attacker_{\it LDE}$, the adversary 
first generates her training data as mentioned in \autoref{section:gap}.
She then trains $\attacker_{\it LDE}$ with the posterior difference $\shadowAV^1 \cdots \shadowAV^m$ 
as the input and the normalized label distribution of their corresponding updating sets as the output.
We assume the adversary knows the cardinality of the updating set.
We try to relax this assumption later in our evaluation.

\smallskip
\noindent\textbf{Experimental Setup.}
We evaluate our label distribution estimation attack using updating set of cardinalities 10 and 100.
For the two different cardinalities, we build attack models as mentioned in the methodology.
All data samples in each updating set for the shadow and target models are sampled uniformly, 
thus each sample (in both training and testing set) for the attack model, 
which corresponds to an updating set, has the same label distribution of the original dataset.
We use a batch size of 64 when updating the models.

For evaluation metrics, we calculate KL-divergence 
for each testing sample (corresponding to an updating set on the target model) 
and report the average result over all testing samples (1,000 in total).
Besides, we also measure the accuracy of predicting the most frequent label over samples in the updating set.
We randomly sample a dataset with the same size as the updating set
and use its samples' label distribution as the baseline, namely Random.

\smallskip
\noindent\textbf{Results.}
We report the result for our label distribution estimation attack in \autoref{fig:lde}.
As shown, $\attacker_{\it LDE}$ achieves a significantly better performance than the Random baseline on all datasets.
For the updating set with 100 data samples on the CIFAR-10 dataset, 
our attack achieves 3 and 2.5 times better accuracy and KL-divergence, respectively, than the Random baseline.
Similarly, for the MNIST and Insta-NY datasets, our attack achieves 1.5 and 4.8 times better accuracy, and 2 and 7.9 times better KL-divergence.
Furthermore, $\attacker_{\it LDE}$ achieves a similar improvement over the Random baseline 
for the updating set of size 10.

Recall that the adversary is assumed to know the cardinality of the updating set in order to train her attack model, 
we further test whether we can relax this assumption.
To this end, we first update the shadow model with 100 samples while updating the target model with 10 samples.
As shown in \autoref{fig:lde_kl10} and \autoref{fig:lde_accuracy10} Transfer 100-10, 
our attack still has a similar performance as the original attack.
However, when the adversary updates her shadow model with 10 data samples while the target model is updated with 100 data samples 
(\autoref{fig:lde_kl100} and \autoref{fig:lde_accuracy100} Transfer 10-100),
our attack performance drops significantly, in particular for KL-divergence on the CIFAR-10 dataset.
We believe this is due to the 10 samples not providing enough information 
for the attack model to generalize to a larger updating set.

\subsection{Multi-sample Reconstruction Attack}
\label{section:MSRAttack}

\noindent\textbf{Attack Definition.}
Our last attack, namely multi-sample reconstruction attack,
aims at reconstructing the updating set.
This attack can be considered as a generalization of the single-sample reconstruction attack, 
and a step towards the goal of reconstructing the training set of a black-box ML model.
Formally, the attack is defined as follows:
\[
\attacker_{\it MSR}: \attackVector \mapsto  \dataset_{\it update}
\]
where $\dataset_{\it update} = \{ \featurevec^{1}_{\it update}, \dots, \featurevec^{\vert \dataset_{\it update} \vert}_{\it update}\}$ contains 
the samples used to update the model.

\smallskip
\noindent\textbf{Methodology.}
The complexity of the task for reconstructing an updating set increases significantly
when the updating set size grows from one to multiple.
Our single-sample reconstruction attack (\autoref{subsection:SSRA}) uses 
AE to reconstruct a single sample.
However, AE cannot generate a set of samples. 
In fact, directly predicting a set of examples is a very challenging task. 
Therefore, we rely on generative models which are able to generate multiple samples rather than a single one.

We first introduce the classical Generative Adversarial Networks (GANs) 
and point out why classical GANs cannot be used for our multi-sample reconstruction attack. 
Next, we propose our Conditional Best of Many GAN ({\UGAN}), 
a novel hybrid generative model and demonstrate how to use it to execute the multi-sample reconstruction attack.

\begin{figure}[!t]
\centering
\includegraphics[width=1\columnwidth]{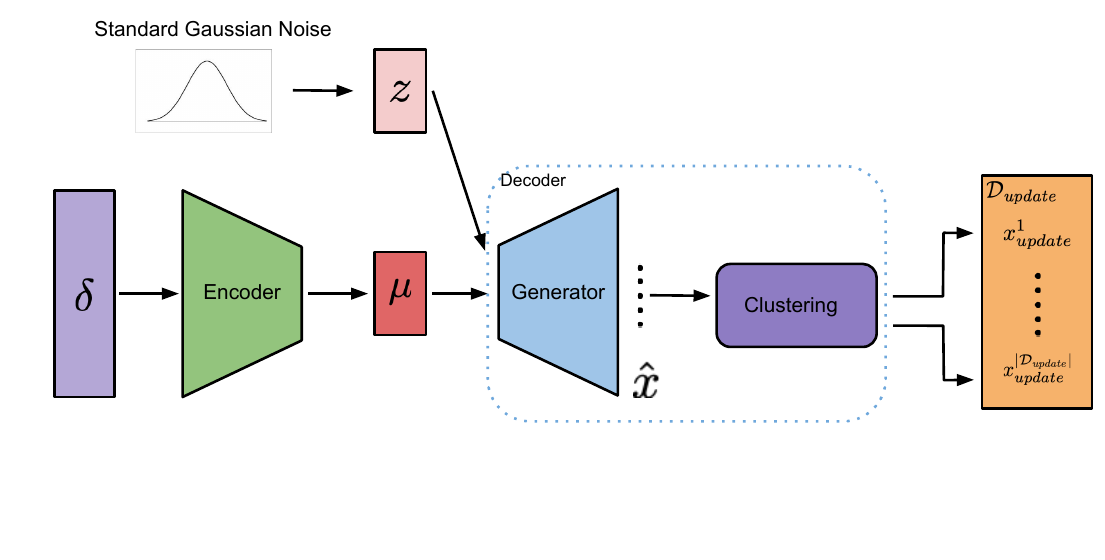}
\caption{Methodology of the multi-sample reconstruction attack ($\attacker_{\it MSR}$).}
\label{fig:msr_overview}
\end{figure}

\begin{figure*}[!t]
\centering
\begin{subfigure}{0.47\columnwidth}
\centering
\includegraphics[width=\columnwidth]{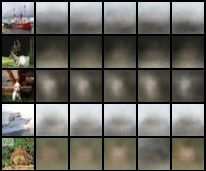}
\caption{}
\label{fig:CIFAR_TOP5_a}
\end{subfigure}
\begin{subfigure}{0.47\columnwidth}
\centering
\includegraphics[width=\columnwidth]{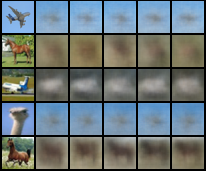}
\caption{}
\label{fig:CIFAR_TOP5_b}
\end{subfigure}
\begin{subfigure}{0.47\columnwidth}
\centering
\includegraphics[width=\columnwidth]{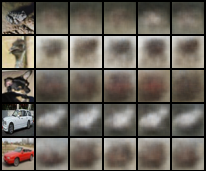}
\caption{}
\label{fig:CIFAR_TOP5_c}
\end{subfigure}
\begin{subfigure}{0.47\columnwidth}
\centering
\includegraphics[width=\columnwidth]{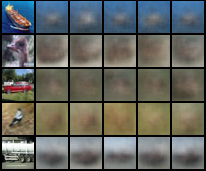}
\caption{}
\label{fig:CIFAR_TOP5_d}
\end{subfigure}
\caption{Visualization of some generated samples from the multi-sample reconstruction attack ($\attacker_{\it MSR}$) 
before clustering on the CIFAR-10 dataset.
Samples are fair random draws, not cherry-picked.
The left column shows the original samples and the next 5 columns show the 5 nearest reconstructed samples
with respect to mean squared error.}
\label{fig:CIFAR_TOP5}
\end{figure*}
 
\smallskip
\noindent\emph{Generative Adversarial Networks.}
Samples from a dataset are essentially samples drawn from a complex data distribution. 
Thus, one way to reconstruct the dataset $\dataset_{\it update}$ 
is to learn this complex data distribution and sample from it.
This is the approach we adopt for our multi-sample reconstruction attack. 
Mainly, the adversary starts the attack by learning the data distribution of $\update$, 
then she generates multiple samples from the learned distribution, 
which is equivalent to reconstructing the dataset $\update$.
In this work, we leverage the state-of-the-art generative model GANs,
which has been demonstrated effective on learning a complex data distribution.

A GAN consists of a pair of ML models: 
a generator (G) and a discriminator (D). The generator G learns to transform 
a Gaussian noise vector $z \sim \mathcal{N}(0,1)$ to a data sample $\hat{x}$, 
\begin{align*}
\text{G}: z \mapsto \hat{x}
\end{align*}
such that the generated sample $\hat{x}$ is indistinguishable from a true data sample. 
This is enabled by the discriminator D which is jointly trained. The generator G tries to fool the discriminator, 
which is trained to distinguish between samples from the Generator (G) and true data samples. 
The objective function maximized by GAN's discriminator D is,
\begin{equation}
    \mathcal{L}_{D} = \mathbb{E}_{x \in \dataset_{\it update}} \log( \text{D}(x) ) + \mathbb{E}_{\hat{x}}\log(1 - \text{D}(\hat{x}))
    \label{eq:discriminator}
\end{equation}
The GAN discriminator D is trained to output 1 (``true'') for real data 
and 0 (``false'') for fake data. On the other hand, the generator G maximizes:
\begin{align*}
    \mathcal{L}_{G} = \mathbb{E}_{\hat{x}} \log( \text{D}(\hat{x}) )
\end{align*}
Thus, G is trained to produce samples $\hat{x} = \text{G}(z)$ that are classified as ``true'' (real) by D.

However, our attack aims to reconstruct $\dataset_{\it update}$ 
for any given $\attackVector$,
which the standard GAN does not support.
Therefore, first, we change the GAN into a conditional model
to condition its generated samples $\hat{x}$ 
on the posterior difference $\attackVector$.
Second, we construct our novel hybrid generative model \UGAN, 
by adding a new ``Best Match'' loss to reconstruct \textit{all} samples inside the updating set \textit{accurately}.

\smallskip
\noindent\emph{{\UGAN}.} 
The decoder of our attack model is casted as our {\UGAN}'s generator (G).
To enable this, we concatenate the noise vector $z$ and the latent vector $\latentVector$ 
produced by our attack model's encoder (with posterior different as input),
and use it as {\UGAN}'s generator's input, as in Conditional GANs~\cite{MO14}.
This allows our decoder to map the posterior difference $\attackVector$ to samples in $\update$. 

However, Conditional GANs are severely prone to mode collapse, 
where the generator's output is restricted to a limited subset of the distribution~\cite{BDS19,YHJZL19}. 
To deal with this, we introduce a reconstruction loss. 
This reconstruction loss forces our GAN to cover all the modes of the distribution (set) of data samples 
used to update the model. 
However, it is unclear, given a posterior difference $\attackVector$ 
and a noise vector $z$ pair, which sample in the data distribution we should force {\UGAN} to reconstruct. 
Therefore, we allow our GAN full flexibility in learning a mapping 
from posterior difference and noise vector $z$ pairs 
to data samples -- this means we allow it to choose the data sample to reconstruct.
We realize this using a novel ``Best Match'' based objective in the {\UGAN} formulation,
\begin{equation}
    \mathcal{L}_{BM} = \sum_{x \in \dataset_{\it update}} \min\limits_{\hat{x} \sim \text{G}} \lVert \hat{x} - x \lVert^{2}_{2} \, + \sum_{\hat{x}} \, \log(\text{D}(\hat{x}))
    \label{eq:generator}
\end{equation}
where $\hat{x} \sim \text{G}$ represents samples produced 
by our {\UGAN} given a latent vector $\latentVector$ and noise sample $z$. 
The first part of the $\mathcal{L}_{BM}$ objective is based on the standard MSE reconstruction loss 
and forces our {\UGAN} to reconstruct all samples in $\dataset_{\it update}$ 
as the error is summed across $x \in \dataset_{\it update}$. 
However, unlike the standard MSE loss, given a data sample $x \in \dataset_{\it update}$, 
the loss is based only on the generated sample $\hat{x}$ which is closest to the data sample $x \in \dataset_{\it update}$. 
This allows {\UGAN} to reconstruct samples in $\dataset_{\it update}$ 
without having an explicit mapping from posterior difference and noise vector $z$ pairs to data samples, 
as only the ``Best Match'' is penalized. 
Finally, the discriminator D ensures that the samples $\hat{x}$ are indistinguishable from the ``true'' samples of $\dataset_{\it update}$.

\begin{figure*}[!t]
\centering
\begin{subfigure}{0.65\columnwidth}
   \includegraphics[width=\columnwidth]{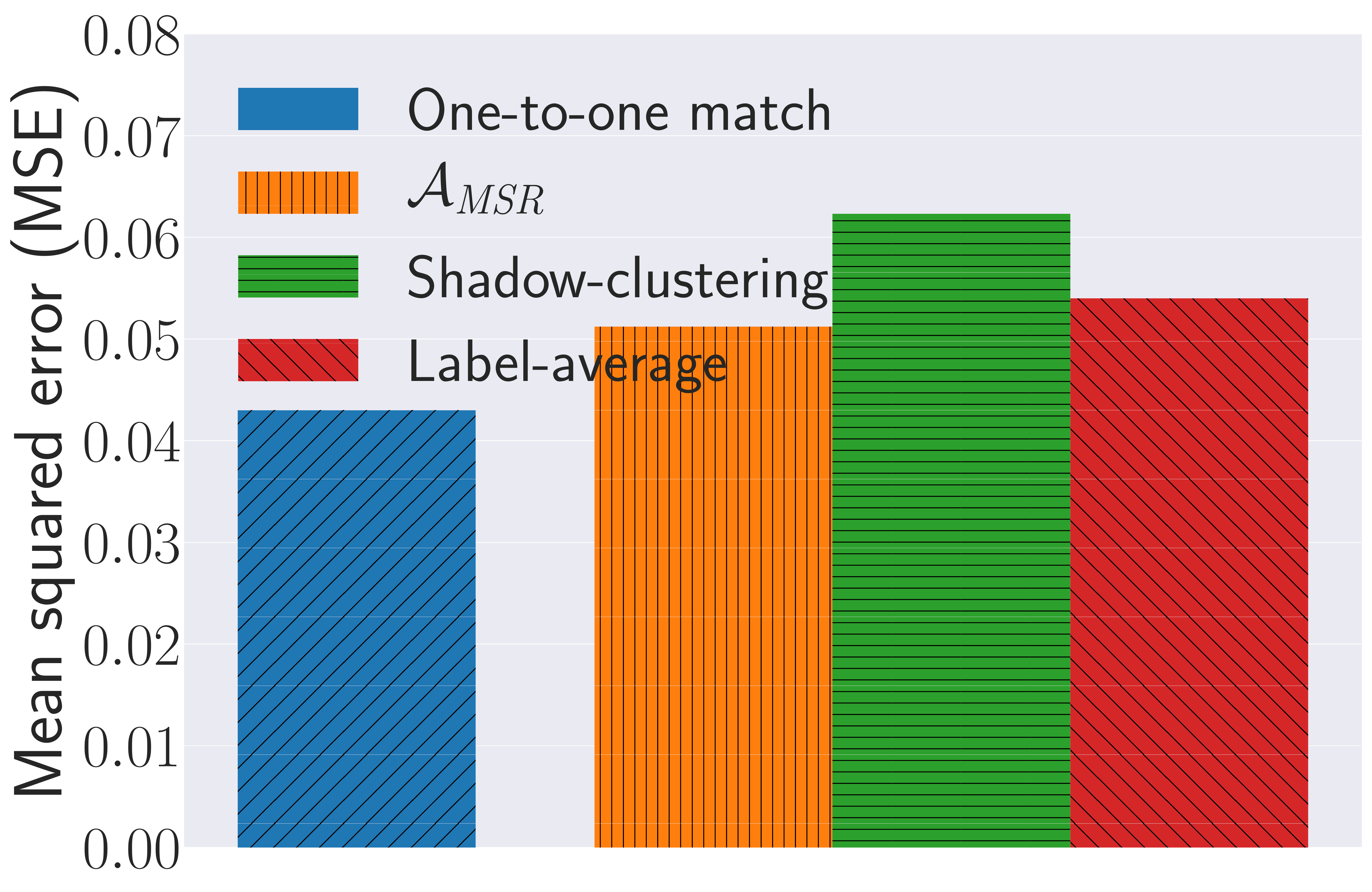}
   \caption{MNIST}
   \label{fig:mse_mnist} 
\end{subfigure}
\begin{subfigure}{0.65\columnwidth}
   \includegraphics[width=\columnwidth]{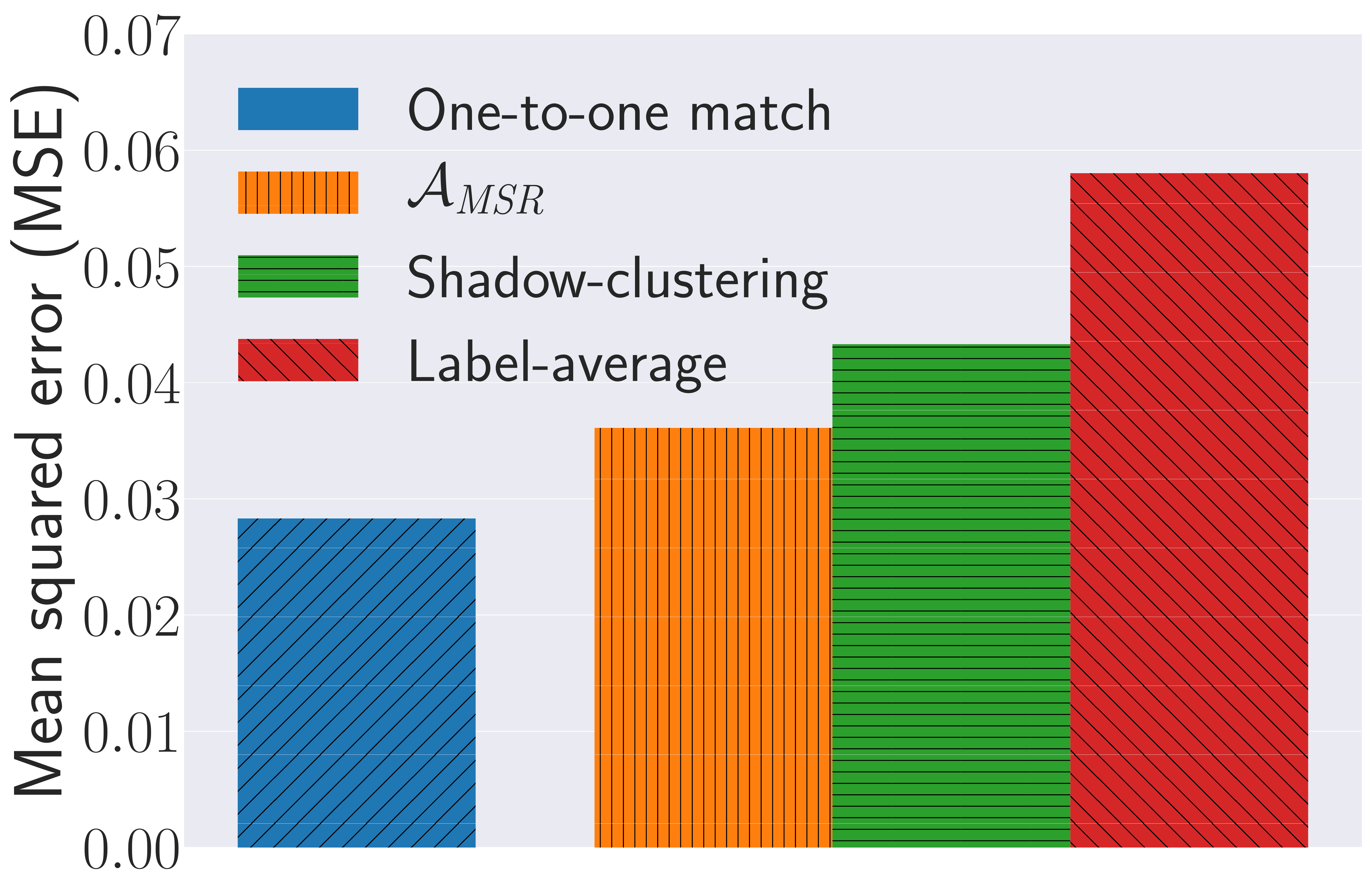}
   \caption{CIFAR-10}
   \label{fig:mse_cifar10} 
\end{subfigure}
\begin{subfigure}{0.65\columnwidth}
   \includegraphics[width=\columnwidth]{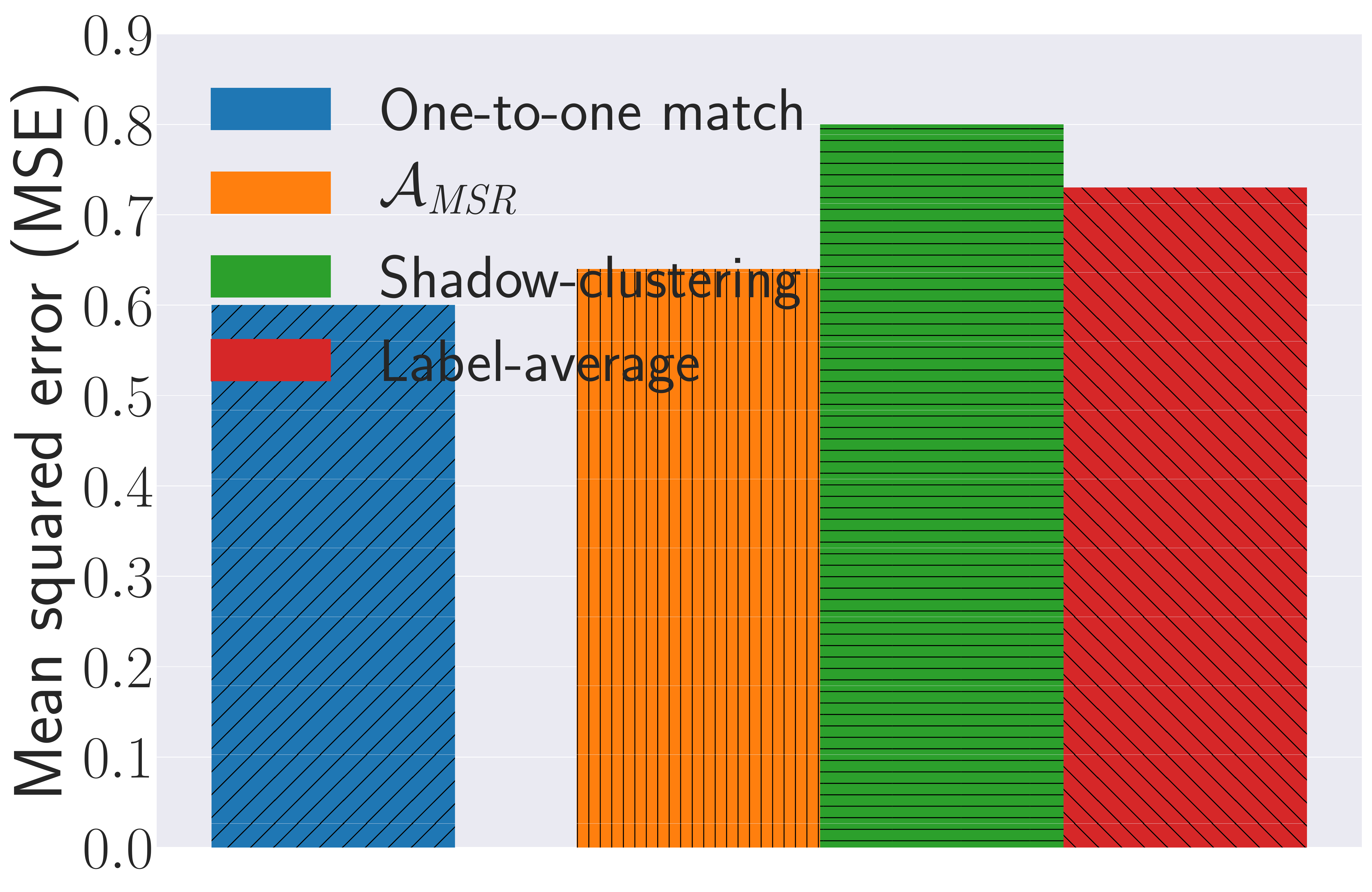}
   \caption{Insta-NY}
   \label{fig:mse_location} 
\end{subfigure}
\caption{[Lower is better] Performance of the multi-sample reconstruction attack ($\attacker_{\it MSR}$) 
together with one-to-one match and the two baseline models.
Mean squared error (MSE) is adopted as the evaluation metric.
The match between the original and reconstructed samples is performed 
by the Hungarian algorithm for both $\attacker_{\it MSR}$ and Shadow-clustering. 
For Label-average, each sample is matched within the average of samples with the same class in the shadow dataset.
One-to-one match serves as an oracle as the adversary cannot use it for her attack.} 
\label{fig:msr_mse} 
\end{figure*}

\smallskip
\noindent\emph{Training of {\UGAN}.}
The training of the attack model $\attacker_{\it MSR}$ is more complicated than previous attacks,
hence we provide more details here.
Similar to the previous attacks, the adversary 
starts the training by generating the training data as mentioned in \autoref{section:gap}.
She then jointly trains her encoder and {\UGAN}\ 
with the posterior difference $\shadowAV^1 \cdots \shadowAV^m$ 
as the inputs and samples inside their corresponding updating sets, 
i.e., $\shadowData^{\it update^1} \cdots \shadowData^{\it update^m}$ as the output.
More concretely, for each posterior difference $\shadowAV^i$, 
she updates her attack model $\attacker_{\it MSR}$ as follows:

\begin{enumerate}
\item The adversary sends the posterior difference $\shadowAV^i$ to her encoder to get the latent vector $\latentVector_i$.
\item She then generates $\vert\shadowData^{\it update^i} \vert $ noise vectors.
\item To create generator's input, she concatenates each of the noise vectors with the latent vector $\latentVector_i$.
\item On the input of the concatenated vectors, the {\UGAN} generates $\vert\shadowData^{\it update^i} \vert $ samples, 
i.e., each vector corresponds to each sample.
\item The adversary then calculates the generator loss as introduced by \autoref{eq:generator}, 
and uses it to update the generator and the encoder.
\item Finally, she calculates and updates the {\UGAN}'s discriminator according to \autoref{eq:discriminator}.
\end{enumerate}

\begin{figure*}[!t]
\centering
\includegraphics[width=0.15\columnwidth]{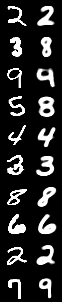}
\includegraphics[width=0.15\columnwidth]{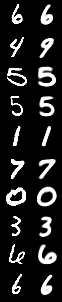}
\includegraphics[width=0.15\columnwidth]{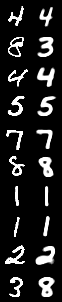}
\includegraphics[width=0.15\columnwidth]{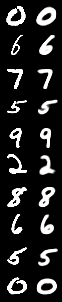}
\includegraphics[width=0.15\columnwidth]{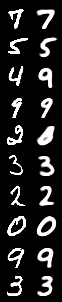}
\includegraphics[width=0.15\columnwidth]{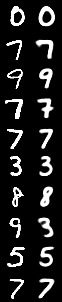}
\includegraphics[width=0.15\columnwidth]{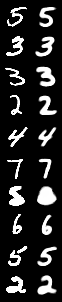}
\includegraphics[width=0.15\columnwidth]{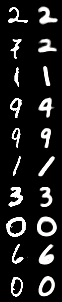}
\includegraphics[width=0.15\columnwidth]{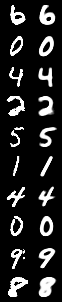}
\includegraphics[width=0.15\columnwidth]{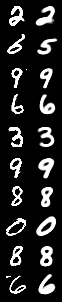}
\caption{
Visualization of a full MNIST updating set together with the output of the multi-sample reconstruction attack ($\attacker_{\it MSR}$) 
after clustering.
Samples are fair random draws, not cherry-picked.
The left column shows the original samples and the right column shows the reconstructed samples.
The match between the original and reconstructed samples is performed by the Hungarian algorithm.
}
\label{fig:MSRA_MNIST}
\end{figure*}

\smallskip
\noindent\emph{Clustering.}
{\UGAN} only provides a generator which learns the distribution of the samples in the updating set.
However, to reconstruct the exact data samples in $\update$,
we need a final step assisted by machine learning clustering.
In detail, we assume the adversary knows the cardinality of $\update$ as in \autoref{section:LDEAttack}.
After {\UGAN} is trained, the adversary utilizes {\UGAN}'s generator to generate a large number of samples.
She then clusters the generated samples into $\vert\update \vert$ clusters.
Here, the K-means algorithm is adopted to perform clustering where we set K to $\vert\update \vert$.
In the end, for each cluster, the adversary calculates its centroid,
and takes the nearest sample to the centroid as one reconstructed sample.

\autoref{fig:msr_overview} presents a schematic view of our multi-sample reconstruction attack's methodology.
The concrete architecture of {\UGAN}'s generator 
and discriminator for the three datasets used in this paper are listed in~\autoref{section:MSRArc}.

\smallskip
\noindent\textbf{Experimental Setup.}
We evaluate the multi-sample reconstruction attack on the updating set of size 100 
and generate 20,000 samples for each updating set reconstruction
with {\UGAN}.
For the rest of the experimental settings, 
we follow the one mentioned in \autoref{section:LDEAttack} except for evaluation metrics and baseline.

We use MSE between the updating and reconstructed data samples 
to measure the multi-sample reconstruction attack's performance.
We construct two baselines, namely Shadow-clustering and Label-average.
For Shadow-clustering, we perform K-means clustering on the adversary's shadow dataset.
More concretely, we cluster the adversary's shadow dataset into 100 clusters
and take the nearest sample to the centroid of each cluster as one reconstructed sample.
For Label-average, we calculate the MSE 
between each sample in the updating set 
and the average of the images with the same label in the adversary's shadow dataset.

\smallskip
\noindent\textbf{Results.}
In \autoref{fig:CIFAR_TOP5},
we first present some visualization of the intermediate result of our attack, 
i.e., the {\UGAN}'s output before clustering, on the CIFAR-10 dataset.
For each randomly sampled image in the updating set, 
we show the 5 nearest reconstructed images with respect to MSE generated by {\UGAN}.
As we can see, our attack model tries to generate images with similar characteristics to the original images.
For instance, the 5 reconstructed images for the airplane image in \autoref{fig:CIFAR_TOP5_b}
all show a blue background and a blurry version of the airplane itself.
The similar result can be observed from the boat image in \autoref{fig:CIFAR_TOP5_a}, 
the car image in \autoref{fig:CIFAR_TOP5_c}, and the boat image in \autoref{fig:CIFAR_TOP5_d}.
It is also interesting to see that {\UGAN} provides different samples 
for the two different horse images in \autoref{fig:CIFAR_TOP5_b}.
The blurriness in the results is expected, 
due to the complex nature of the CIFAR-10 dataset and the weak assumptions for our adversary, 
i.e., access to black-box ML model.

We also quantitatively measure the performance of our intermediate results, 
by calculating the MSE between each image in the updating set and its nearest reconstructed sample.
We refer to this as one-to-one match.
\autoref{fig:msr_mse} shows for the CIFAR-10, MNIST, and Insta-NY datasets, 
we achieve 0.0283, 0.043 and 0.60 MSE, respectively.
It is important to note that the adversary cannot perform one-to-one match 
as she does not have access to ground truth samples in the updating set,
i.e., one-to-one match is an oracle.

\autoref{fig:msr_mse} shows the mean squared error of our full attack with clustering for all datasets.
To match each of our reconstructed samples to a sample in $\update$,
we rely on the Hungarian algorithm~\cite{K55}.
This guarantees that each reconstructed sample is only matched with one ground truth sample in $\update$ and vice versa.
As we can see, our attack outperforms 
both baseline models on the CIFAR-10, MNIST and Insta-NY datasets
(20\%, 22\%, and 25\% performance gain 
for Shadow-clustering and 60.1\%, 5.5\% and 14\% performance gain for Label-average, respectively).
The different performance gain of our attack over the label-average baseline 
for different datasets is due to the different complexity of these datasets. 
For instance, all images inside MNIST have black background 
and lower variance within each class compared to the CIFAR-10 dataset. 
The different complexity results in some datasets having a more representative label-average, 
which leads to a lower performance gain of our attack over them.

These results show that our multi-sample reconstruction attack provides 
a more useful output than calculating the average from the adversary's dataset.
In detail, our attack achieves an MSE of 0.036 on the CIFAR-10 dataset, 
0.051 on the MNIST dataset, and 0.64 on the Insta-NY dataset.
As expected, the MSE of our final attack is higher than one-to-one match, 
i.e., the above mentioned intermediate results.

We further visualize our full attack's result on the MNIST dataset.
\autoref{fig:MSRA_MNIST} shows a sample of a full MNIST updating set reconstruction, 
i.e., the {\UGAN}'s reconstructed images for the 100 original images in an updating set.
We observe that our attack model reconstructs diverse digits of each class 
that for most of the cases match the actual ground truth data very well.
This suggests {\UGAN} is able to capture most modes in a data distribution well.
Moreover, comparing the results of this attack (\autoref{fig:MSRA_MNIST}) 
with the results of the single-sample reconstruction attack (\autoref{fig:singlePointMNIST}), 
we can see that this attack produces sharper images. 
This result is due to the discriminator of our {\UGAN}, 
as it is responsible for making the {\UGAN}'s output to look real, i.e., sharper in this case.

One limitation of our attack is that {\UGAN}'s sample generation and clustering are performed separately.
In the future, we plan to combine them to perform an end-to-end training 
which may further boost our attack's performance.

From all these results, we show that our attack 
does not generate a general representation of data samples affiliated with the same label, 
but tries to reconstruct images with similar characteristics 
as the images inside the updating set 
(as shown by the different shapes of the same numbers in~\autoref{fig:MSRA_MNIST}).

\smallskip
\noindent\textbf{Relaxing The Knowledge of Updating Set Cardinality.}
One of the above attack’s main assumptions is the adversary's knowledge of
the updating set cardinality, i.e., $|\update|$. 
Next, we show how to relax this assumption. 
To recap, the adversary needs the updating set cardinality 
when updating her shadow model and clustering {\UGAN}'s output. 
We address the former by using updating sets of different cardinalities. 
For the latter, we use the silhouette score to find the optimal k for K-means, 
i.e., the most likely value of the target updating set's cardinality. 
The silhouette score lies in the range between \mbox{-1} and 1, it reflects the consistency of the clustering. 
Higher silhouette score leads to more suitable k.

Specifically, the adversary follows the previously presented methodology 
in \autoref{section:MSRAttack} with the following modifications. 
First, instead of using updating sets with the same cardinality, the adversary uses updating sets with different cardinalities
to update the shadow model.
Second, after the adversary generates multiple samples from {\UGAN}, she uses the silhouette score to find the optimal k.
The silhouette score is used here to identify the target model's updating set cardinality 
from the different updating sets cardinalities used to update the shadow model.

We evaluate the effectiveness of this attack on all datasets. 
We use a target model updated with 100 samples 
and create our shadow updated models using updating sets with cardinality 10 and 100. 
Concretely, we update the shadow model half of the time 
with updating sets of cardinality 10 and the other half with cardinality 100.

Our evaluation shows that our attack consistently 
produces higher silhouette score -by at least 20\%- for the correct cardinality in all cases.
In another way, our method can always detect the right cardinality of the updating set in this setting.
Moreover, the MSE for the final output of the attack 
only drops by 1.6\%, 0.8\%, and 5.6\% for the Insta-NY, MNIST, and CIFAR-10 datasets, respectively.
\section{Discussion}
\label{section:further_analysis}

In this section, we analyze the effect of different hyperparameters of both the target and shadow models 
on our attacks' performance. 
Furthermore, we investigate relaxing the threat model assumptions and discuss the limitations of our attacks.

\smallskip
\noindent\textbf{Relaxing The Attacker Model Assumption.}
Our threat model has two main assumptions:
Same data distribution for both target and shadow datasets 
and same structure for both target and shadow models.
We relax the former by proposing data transferability attack
and latter by model transferability attack.

\begin{table}[!t]
\centering
\begin{tabular}{r|c|c} 
\toprule
Attack& Original&Transfer\\
\midrule
$\attacker_{\it LI}$& 0.97  & 0.89\\
$\attacker_{\it SSR}$& 0.68 &1.1\\
$\attacker_{\it LDE}(10)$& 0.59(0.0317) & 0.55(0.0377) \\
$\attacker_{\it LDE}(100)$& 0.89(0.0041)  & 0.89 (0.0067)\\
$\attacker_{\it MSR}$& 0.64  & 0.73\\
\bottomrule
\end{tabular}
\caption{Evaluation of the data transferability attacks.
The first column shows all different attacks, 
the second and third shows the performance of the attacks 
using similar and different distributions, respectively.
Where $\attacker_{\it LI}$ performance is measured in accuracy, 
$\attacker_{\it MSR}$ and $\attacker_{\it SSR}$ measured in MSE, 
and $\attacker_{\it LDE}(10)$ and $\attacker_{\it LDE}(100)$ measured in accuracy (KL-divergence).}
\label{table:DataTransferablity}
\end{table}

\smallskip
\noindent\emph{Data Transferability.}
In this setting,
we locally train and update the shadow model with a dataset which comes from a different distribution
from the target dataset.
For our experiments,
we use Insta-NY as the target dataset and Insta-LA as the shadow dataset.

\autoref{table:DataTransferablity} depicts the evaluation results.
As expected, the performance of our data transferability attacks drops; 
however, they are still significantly better than corresponding baseline models. 
For instance, the performance of the multi-sample reconstruction attack drops by 14\% 
but is still 10\% better than the baseline (see~\autoref{fig:msr_mse}). 
Moreover, the multi-sample label distribution attack's accuracy (KL-divergence) 
only drops by 6.8\% (18.9\%) and 0\% (63\%), 
which is still significantly better than the baseline (see~\autoref{fig:lde}) 
by 6.5x (2x) and 4.6x (4.8x) for updating set sizes of 10 and 100, respectively.

\smallskip
\noindent\emph{Model Transferablity.}
Now we relax the attacker's knowledge on the target model's architecture, 
i.e., we use different architectures for shadow and target models. 
In our experiments on Insta-NY,
we use the same architecture mentioned previously 
in~\autoref{subsection:LPAttack} for the target model, 
and remove one hidden layer and use half of the number of neurons 
in other hidden layers for the shadow model.
 
The performance drop of our model transferability attack 
is only less than 2\% for all of our attacks, 
which shows that our attacks are robust against such changes in the model architectures.
We observe similar results when repeating the experiment using different architectures
and omit them for space restrictions.

\smallskip
\noindent\textbf{Effect of The Probing Set Cardinality.}
We evaluate the performance of our attacks on CIFAR-10 
when the probing set cardinality is 10, 100, 1,000, or 10,000.
As our encoder's input size relies on the probing set cardinality (see~\autoref{section:gap}),
we adjust its input layer size accordingly.

As expected, using a probing set of size 10 reduces the performance of the attacks.
For instance, the single-sample label inference 
and reconstruction attacks' performance drops by 9\% and 71\%, respectively.
However, increasing the probing set cardinality from 100 to 1,000 or 10,000 
has a limited effect (up to 3.5\% performance gain).
It is also important to mention that the computational requirement 
for our attacks increases with an increasing probing set cardinality,
as the cardinality decides the size of the input layer for our attack models.
In conclusion, using 100 samples for probing the target model is a suitable choice.

\smallskip
\noindent\textbf{Effect of Target Model Hyperparameters.}
We now evaluate our attacks' performance with respect to two hyperparameters of the target model. 

\smallskip
\noindent\emph{Target Model's Training Epochs Before Updating.}
We use the MNIST dataset to evaluate the multi-sample 
label distribution estimation attack's performance on target models trained for 10, 20, and 50 epochs. 
For each setting, we update the model and execute our attack as mentioned in~\autoref{section:LDEAttack}.

The experiments show that the difference in the attack's performance for the different models is less than 2\%. 
That is expected as gradients are not monotonically decreasing during the training procedure. 
In other words, information is not necessarily vanishing~\cite{GBC16}.

\smallskip
\noindent\emph{Target Model's Updating Epochs.}
We train target and shadow models as introduced in~\autoref{section:LDEAttack} 
with the Insta-NY dataset, but we update the models using different number of epochs. 
More concretely, we update the models using from 2 to 10 epochs 
and evaluate the multi-sample label distribution estimation attack's performance on the updated models.

\begin{figure}[!t]
\centering
\includegraphics[width=0.8\columnwidth]{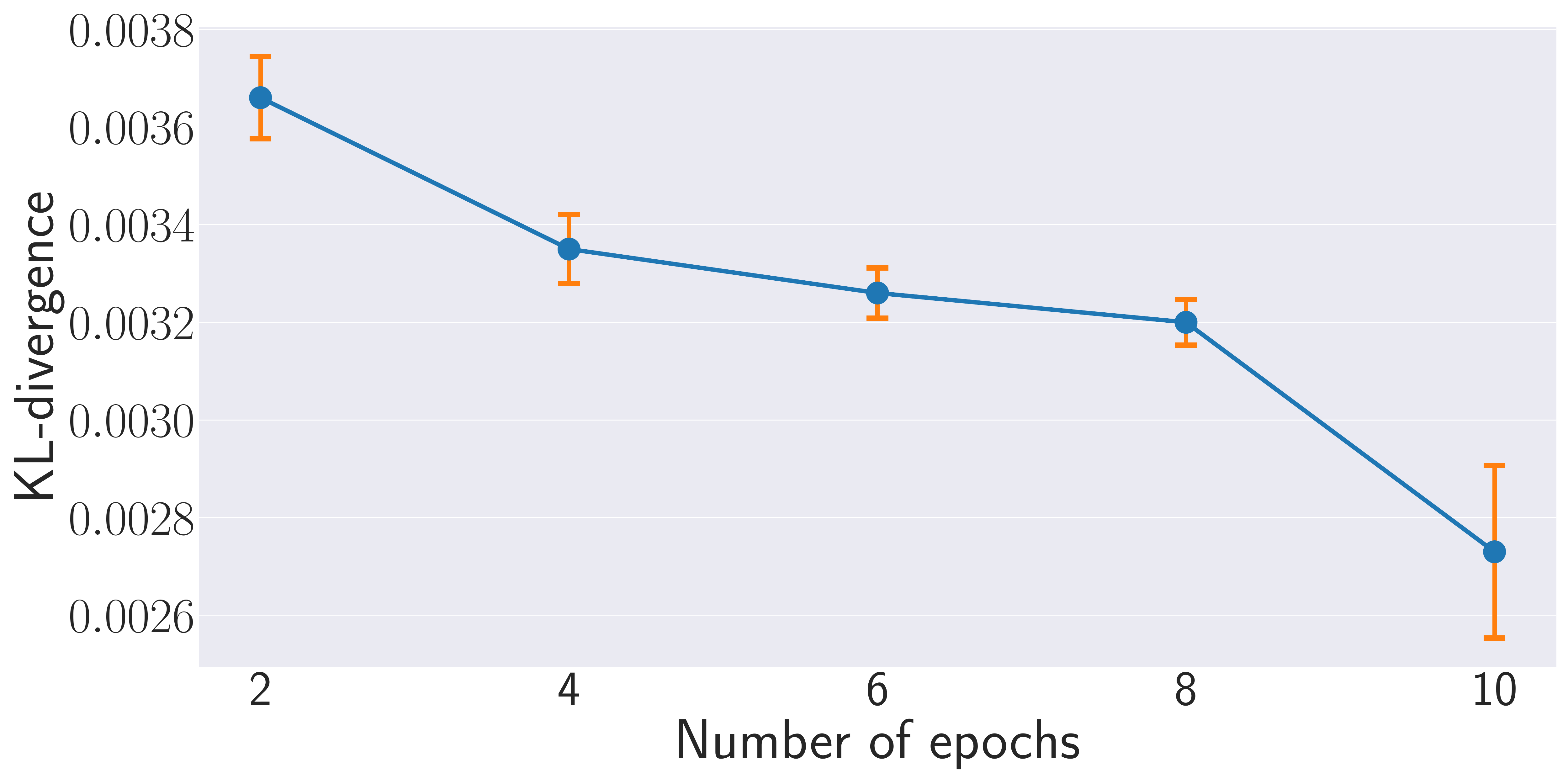}
\caption{[Lower is better] The performance of the multi-sample label distribution estimation attack ($\attacker_{\it LDE}$) 
with different number of epochs used to update the target model. 
}
\label{fig:updatingEpochs_kl}
\end{figure}

We report the results of our experiments in~\autoref{fig:updatingEpochs_kl}. 
As expected, the multi-sample label distribution estimation attack's performance 
improves with the increase of the number of epochs used to update the model. 
For instance, the attack performance improves by 25.4 \% 
when increasing the number of epochs used to update the model from 2 to 10.

\smallskip
\noindent\textbf{Limitations of Our Attacks.}
For all of our attacks, we assume a simplified setting, 
in which, the target model is solely updated on new data. 
Moreover, we perform our attacks on updating sets of maximum cardinality of 100. 
In future work, we plan to further investigate a more complex setting, 
where the target model is updated using larger updating sets of both new and old data.
\section{Possible Defenses}
\label{section:defense}

\noindent\textbf{Adding Noise to Posteriors.}
All our attacks leverage posterior difference as the input.
Therefore, to reduce our attacks' performance, one could sanitize posterior difference.
However, the model owner cannot directly manipulate the posterior difference, 
as she does not know with what or when the adversary probes her model.
Therefore, she has to add noise to the posterior for each queried sample independently.
We have tried adding noise sampled from a uniform distribution to the posteriors.
Experimental results show that the performance for some of our attacks indeed drops to a certain degree.
For instance, the single-sample label inference attack on the CIFAR-10 dataset drops by 17\% in accuracy.
However, the performance of our multi-sample reconstruction attack stays stable.
One reason might be the noise vector $z$ is part of \UGAN 's input which makes the attack model more robust to the noisy input.

\smallskip
\noindent\textbf{Differential Privacy.}
Another possible defense mechanism against our attacks is differentially private learning.
Differential privacy~\cite{DR14} can help an ML model learn its main tasks while reducing its memory on the training data.
If differentially private learning schemes~\cite{ACGMMTZ16,SS15,CM09} are used when updating the target ML model,
this by design will reduce the performance of our attacks.
However, it is also important to mention that depending on the privacy budget for differential privacy, 
the utility of the model can drop significantly.

\smallskip
We leave an in-depth exploration of effective defense mechanisms against our attacks as a future work.
\section{Related Works}
\label{section:related}

\noindent\textbf{Membership Inference.}
Membership inference aims at determining whether a data sample is inside a dataset.
It has been successfully performed in various settings, 
such as biomedical data~\cite{HSRDTMPSNC08,HZHBTWB19} and location data~\cite{PTC18,PTC19}.
Shokri et al.~\cite{SSSS17} propose the first membership inference attack against machine learning models.
In this attack, an adversary's goal is to determine whether a data sample is 
in the training set of a black-box ML model.
To mount this attack, the adversary relies on a binary machine learning classifier
which is trained with the data derived from shadow models (similar to our attacks).
More recently, multiple membership inference attacks have been proposed 
with new attacking techniques or targeting 
on different types of ML models~\cite{LBG17,HMDC19,LBWBWTGC18,YGFJ18,NSH18,SS18,SZHBFB19,NSH19}.

In theory, membership inference attack can be used to reconstruct the dataset,
similar to our reconstruction attacks.
However, it is not scalable in the real-world setting
as the adversary needs to obtain a large-scale dataset 
which includes all samples in the target model's training set.
Though our two reconstruction attacks are designed specifically for the online learning setting,
we believe the underlying techniques we propose, 
i.e., pretrained decoder from a standard autoencoder and {\UGAN},
can be further extended to reconstruct datasets from black-box ML models
in other settings.

\smallskip
\noindent\textbf{Model Inversion.}
Fredrikson et al.~\cite{FLJLPR14} propose model inversion attack first on biomedical data.
The goal of model inversion is to infer some missing attributes 
of an input feature vector based on the interaction with a trained ML model.
Later, other works generalize the model inversion attack to other settings, 
e.g.,, reconstructing recognizable human faces~\cite{FJR15,HAP17}.
As pointed out by other works~\cite{SSSS17,MSCS19}, 
model inversion attack reconstructs a general representation of data samples
affiliated with certain labels,
while our reconstruction attacks target on specific data samples used in the updating set.

\smallskip
\noindent\textbf{Model Stealing.}
Another related line of work is model stealing.
Tram{\`{e}}r et al.~\cite{TZJRR16} are among the first 
to introduce the model stealing attack against black-box ML models.
In this attack, an adversary tries to learn the target ML model's parameters.
Tram{\`{e}}r et al.\ propose various attacking techniques including equation-solving and decision tree path-finding.
The former has been demonstrated to be effective on simple ML models, such as logistic regression,
while the latter is designed specifically for decision trees, a class of machine learning classifiers.
Moreover, relying on an active learning based retraining strategy,
the authors show that it is possible to steal an ML model 
even if the model only provides the label instead of posteriors as the output.
More recently, Orekondy et al.~\cite{OSF19} propose a more advanced attack 
on stealing the target model's functionality
and show that their attack is able to replicate a mature commercial machine learning API.
In addition to model parameters, 
several works concentrate on stealing ML models' hyperparameters~\cite{OASF18,WG18}.

\smallskip
Besides the above, there exist a wide range of other attacks and defenses on machine learning 
models~\cite{CM09,SZSBEGF13,VL14,GSS15,LV15,ACGMMTZ16,PMGJCS17,GSBRDZE17,SRS17,
CW17,TKPGBM17,YVCZZ17,XEQ18,JOBLNL18,GMXSX18,ZHRLPB18,ACW18,GWYGB18,WYVZZ18,ZE19,YKT19,JSBZG19,LHZG19,ZHSMVB20}
\section{Conclusion}
\label{section:conclusion}

Large-scale data being generated at every second
turns ML model training into a continuous process.
In consequence, a machine learning model queried 
with the same set of data samples at two different time points 
will provide different results.
In this paper, we investigate whether these different model outputs can constitute a new attack surface
for an adversary to infer information of the dataset used to perform model update.
We propose four different attacks in this surface
all of which follow a general encoder-decoder structure.
The encoder encodes the difference in the target model's output before and after being updated, 
and the decoder generates different types of information regarding the updating set.

We start by exploring a simplified case when an ML model is only updated with one single data sample.
We propose two different attacks for this setting.
The first attack shows that the label of the single updating sample can be effectively inferred.
The second attack utilizes an autoencoder's decoder as the attack model's pretrained decoder
for single-sample reconstruction.

We then generalize our attacks to the case when the updating set contains multiple samples.
Our multi-sample label distribution estimation attack trained following a KL-divergence loss
is able to infer the label distribution of the updating set's data samples effectively.
For the multi-sample reconstruction attack,
we propose a novel hybrid generative model, namely {\UGAN}, which uses a ``Best Match' loss in its objective function.
The ``Best Match'' loss directs {\UGAN}'s generator to reconstruct each sample in the updating set.
Quantitative and qualitative results show that our attacks achieve promising performance.

%-------------------------------------------------------------------------------
\section*{Acknowledgments}
%-------------------------------------------------------------------------------

We thank the anonymous reviewers, and our shepherd, David Evans, for their helpful feedback and guidance.

The research leading to these results has received funding from the European Research Council 
under the European Union's Seventh Framework Programme (FP7/2007-2013)/ ERC grant agreement no. 610150-imPACT.
%-------------------------------------------------------------------------------

\bibliographystyle{plain}
\bibliography{normal_generated}

\appendix

\noindent {\Huge \textbf{Appendices}}

\section{Target Models Architecture}
\label{section:targetArc}

\begin{tcolorbox}[boxsep=1pt,left=2pt,right=2pt,top=0.5 pt,bottom=0pt]
\emph{MNIST model:}
\begin{align*}
\texttt{Sample} \rightarrow
\texttt{conv2d(5, 10)} & \\
\texttt{max($2$)} & \\
\texttt{conv2d(5, 20)} & \\
\texttt{max($2$)} & \\
\texttt{FullyConnected(50)} & \\
\texttt{FullyConnected(10)} & \\
\texttt{Softmax}  & \rightarrow \texttt{$\labelVec$}
\end{align*}
\end{tcolorbox}

\begin{tcolorbox}[boxsep=1pt,left=2pt,right=2pt,top=0.5 pt,bottom=0pt]
\emph{CIFAR-10 model:}
\begin{align*}
\texttt{Sample} \rightarrow
\texttt{conv2d(5, 6)} & \\
\texttt{max($2$)} & \\
\texttt{conv2d(5, 16)} & \\
\texttt{max($2$)} & \\
\texttt{FullyConnected(120)} & \\
\texttt{FullyConnected(84)} & \\
\texttt{FullyConnected(10)} & \\
\texttt{Softmax}  & \rightarrow \texttt{$\labelVec$}
\end{align*}
\end{tcolorbox}

\begin{tcolorbox}[boxsep=1pt,left=2pt,right=2pt,top=0.5 pt,bottom=0pt]
\emph{Insta-NY Model:}
\begin{align*}
\texttt{Sample} \rightarrow
\texttt{FullyConnected(32)} & \\
\texttt{FullyConnected(16)} & \\
\texttt{FullyConnected(9)} & \\
\texttt{Softmax}  & \rightarrow \texttt{$\labelVec$}
\end{align*}
\end{tcolorbox}
\noindent Here, \texttt{max($2$)} 
denotes a max-pooling layer with a $2 \times 2$ kernel, 
\texttt{FullyConnected($x$)} denotes a fully connected layer with $x$ hidden units, 
\texttt{Conv2d(k',s')} denotes a 2-dimension convolution layer 
with kernel size $k' \times k'$ and $s'$ filters, and \texttt{Softmax} denotes the Softmax function. 
We adopt ReLU as the activation function for all layers for the MNIST, CIFAR-10 and Location models. 

\section{Encoder Architecture}
\label{section:encArc}

\begin{tcolorbox}[boxsep=1pt,left=2pt,right=2pt,top=0.5 pt,bottom=0pt]
\emph{Encoder architecture}:
\begin{align*}
\texttt{$\attackVector$} \rightarrow 
\texttt{FullyConnected($128$)} & \\
\texttt{FullyConnected($64$)}  & \rightarrow \texttt{$\latentVector$}
\end{align*}
\end{tcolorbox}
\noindent Here, $\latentVector$ denotes the latent vector which serves as the input for our decoder.
Furthermore, we use LeakyReLU as our encoder's activation function 
and apply dropout on both layers for regularization. 

\section{Single-sample Label Inference Attack's Decoder Architecture}
\label{section:SLIArc}

\begin{tcolorbox}[boxsep=1pt,left=2pt,right=2pt,top=0.5 pt,bottom=0pt]
\emph{$\attacker_{\it LI}$'s decoder architecture:}
\begin{align*}
\texttt{$\latentVector$} \rightarrow 
\texttt{FullyConnected($n$)} & \\
\texttt{Softmax}  & \rightarrow \texttt{$\labelVec$}
\end{align*}
\end{tcolorbox}
\noindent Here, $n$ is equal to the size of $\labelVec$, i.e., $n = |\labelVec|$.

\section{Single-sample Reconstruction Attack}
\label{section:SSRArc}

\subsection{AE's Encoder Architecture}

\begin{tcolorbox}[boxsep=1pt,left=2pt,right=2pt,top=0.5 pt,bottom=0pt]
\emph{AE's encoder architecture for MNIST and CIFAR-10:}
\begin{align*}
\texttt{Sample} \rightarrow
\texttt{conv2d($k_1$, $s_1$)} & \\
\texttt{max($2$)} & \\
\texttt{conv2d($k_2$, $s_2$)} & \\
\texttt{max($2$)} & \\
\texttt{FullyConnected($f_1$)} & \\
\texttt{FullyConnected($f_2$)} & \rightarrow \latentVector_{\it AE}
\end{align*}
\end{tcolorbox}

\begin{tcolorbox}[boxsep=1pt,left=2pt,right=2pt,top=0.5 pt,bottom=0pt]
\emph{AE's encoder architecture for Insta-NY:}
\begin{align*}
\texttt{Sample} \rightarrow
\texttt{FullyConnected(64)} & \\
\texttt{FullyConnected(32)} & \\
\texttt{FullyConnected(16)} & \\
\texttt{FullyConnected(16)} & \rightarrow \latentVector_{\it AE}
\end{align*}
\end{tcolorbox}
\noindent Here, $\latentVector_{\it AE}$ is the latent vector output of the encoder. 
Moreover, $k_i$, $s_i$, and $f_i$ represent the kernel size, number of filters, 
and number of units in the $i$th layer. 
The concrete values of these hyperparameters depend on the target dataset, we present our used values in~\autoref{table:AEValues}.
We adopt ReLU as the activation function for all layers for the MNIST and CIFAR-10 encoders. 
For the Insta-NY decoder, we use ELU as the activation function for all layers except for the last one.
Finally, we apply dropout after the first fully connected layer for MNIST and CIFAR-10.
For Insta-NY,
we apply dropout and batch normalization for the first three fully connected layers.

\subsection{AE's Decoder Architecture}

\begin{tcolorbox}[boxsep=1pt,left=2pt,right=2pt,top=0.5 pt,bottom=0pt]
\emph{Autoencoder's decoder architecture  for MNIST and CIFAR-10:}
\begin{align*}
\latentVector_{\it AE} \rightarrow
\texttt{FullyConnected($f_1'$)} & \\
\texttt{FullyConnected($f_2'$)} & \\
\texttt{ConvTranspose2d($k_1'$, $s_1'$)} & \\
\texttt{ConvTranspose2d($k_2'$, $s_2'$)} & \\
\texttt{ConvTranspose2d($k_3'$, $s_3'$)} & \rightarrow \texttt{Sample}
\end{align*}
\end{tcolorbox}

\begin{tcolorbox}[boxsep=1pt,left=2pt,right=2pt,top=0.5 pt,bottom=0pt]
\emph{Autoencoder's decoder architecture  for Insta-NY:}
\begin{align*}
\latentVector_{\it AE} \rightarrow
\texttt{FullyConnected(16)} & \\
\texttt{FullyConnected(32)} & \\
\texttt{FullyConnected(64)} & \\
\texttt{FullyConnected(168)} & \rightarrow \texttt{Sample}
\end{align*}
\end{tcolorbox}
\noindent Here, \texttt{ConvTranspose2d(k',s')} denotes a 2-dimension transposed convolution layer 
with kernel size $k' \times k'$ and $s'$ filters, 
and $f_i'$ specifies the number of units in the $i$th fully connected layer. 
The concrete values of these hyperparameters are presented in~\autoref{table:AEValues}.
For MNIST and CIFAR-10 decoders,
we again use ReLU as the activation function for all layers except for the last one 
where we adopt \texttt{tanh}.
For the Insta-NY decoder, we adopt ELU for all layers except for the last one.
We also apply dropout after the last fully connected layer for regularization for MNIST and CIFAR-10, 
and dropout and batch normalization on the first three fully connected layers for Insta-NY.

\begin{table}
\centering
\caption{Hyperparameters for AE's encoder and decoder.}
\label{table:AEValues}
\begin{tabular}{rcc} 
\toprule
Variable & MNIST & CIFAR-10\\
\midrule
$k_1$ & 3 &3 \\
$s_1$ & 16 &32\\
$k_2$ & 3 &3\\
$s_2$ & 8 & 16\\
$f_1$ & 15 & 50\\
$f_2$ & 10 & 30\\
$f_1'$ &  15& 50\\
$f_2'$ & 32 & 64\\
$k_1'$ & 3 & 3 \\
$s_1'$ & 16  & 32\\
$k_2'$ & 5 & 5 \\
$s_2'$ & 8 & 16\\
$k_3'$ & 2 & 4\\
$s_3'$ & 1 & 3\\
\bottomrule
\end{tabular}
\end{table}

\section{Multi-sample Reconstruction Attack's Decoder Architecture}
\label{section:MSRArc}

\begin{tcolorbox}[boxsep=1pt,left=2pt,right=2pt,top=0.5 pt,bottom=0pt]
\emph{{\UGAN}'s generator architecture for MNIST:}
\begin{align*}
\latentVector, z \rightarrow
\texttt{FullyConnected(2048)} & \\
\texttt{FullyConnected(2048)} & \\
\texttt{FullyConnected(2048)} & \\
\texttt{FullyConnected(784)} & \rightarrow \texttt{Sample}
\end{align*}
\end{tcolorbox}

\begin{tcolorbox}[boxsep=1pt,left=2pt,right=2pt,top=0.5 pt,bottom=0pt]
\emph{{\UGAN}'s discriminator architecture for MNIST:}
\begin{align*}
\latentVector, z \rightarrow
\texttt{FullyConnected(1024)} & \\
\texttt{FullyConnected(512)} & \\
\texttt{FullyConnected(256)} & \\
\texttt{FullyConnected(1)} &\\
\texttt{Sigmoid}\rightarrow &\texttt{\{1,0\}}
\end{align*}
\end{tcolorbox}

\begin{tcolorbox}[boxsep=1pt,left=2pt,right=2pt,top=0.5 pt,bottom=0pt]
\emph{{\UGAN}'s generator architecture for CIFAR-10:}
\begin{align*}
\latentVector, z \rightarrow
\texttt{conv2d(2, 512)} & \\
\texttt{conv2d(4, 256)} & \\
\texttt{conv2d(4, 128)} & \\
\texttt{conv2d(4, 64)} & \\
\texttt{conv2d(4, 3)} & \rightarrow \texttt{Sample}
\end{align*}
\end{tcolorbox}

\begin{tcolorbox}[boxsep=1pt,left=2pt,right=2pt,top=0.5 pt,bottom=0pt]
\emph{{\UGAN}'s discriminator architecture for CIFAR-10:}
\begin{align*}
\latentVector, z \rightarrow
\texttt{conv2d(2, 64)} & \\
\texttt{conv2d(4, 128)} & \\
\texttt{conv2d(4, 256)} & \\
\texttt{conv2d(4, 512)} & \\
\texttt{conv2d(4, 1)} & \\
\texttt{Sigmoid}\rightarrow &\texttt{\{1,0\}}
\end{align*}
\end{tcolorbox}

\begin{tcolorbox}[boxsep=1pt,left=2pt,right=2pt,top=0.5 pt,bottom=0pt]
\emph{{\UGAN}'s generator architecture for Insta-NY:}
\begin{align*}
\latentVector, z \rightarrow
\texttt{FullyConnected(512)} & \\
\texttt{FullyConnected(512)} & \\
\texttt{FullyConnected(256)} & \\
\texttt{FullyConnected(168)} & \rightarrow \texttt{Sample}
\end{align*}
\end{tcolorbox}

\begin{tcolorbox}[boxsep=1pt,left=2pt,right=2pt,top=0.5 pt,bottom=0pt]
\emph{{\UGAN}'s discriminator architecture for Insta-NY:}
\begin{align*}
\latentVector, z \rightarrow
\texttt{FullyConnected(512)} & \\
\texttt{FullyConnected(256)} & \\
\texttt{FullyConnected(128)} & \\
\texttt{FullyConnected(1)} &\\
\texttt{Sigmoid}\rightarrow &\texttt{\{1,0\}}
\end{align*}
\end{tcolorbox}
\noindent Here, for both generators and discriminators, 
\texttt{Sigmoid} is the Sigmoid function, 
batch normalization is applied on the output of each layer except the last layer, 
and LeakyReLU is used as the activation function for all layers except the last one, which uses \texttt{tanh}.
\end{document}